\documentclass[12pt]{article}

\usepackage{amsmath,amsfonts,amssymb,mathrsfs,graphicx,cite}

\newcommand{\capdef}{}
\newcommand{\mycaption}[2][\capdef]{\renewcommand{\capdef}{#2}%
        \caption[#1]{{\footnotesize #2}}}
\makeatletter
\renewcommand{\fnum@table}{\textbf{\tablename~\thetable}}
\renewcommand{\fnum@figure}{\textbf{\figurename~\thefigure}}
\makeatother

\newcounter{myenumi}

\renewcommand{\themyenumi}{\roman{myenumi}}
{\end{list}}%

\newlength{\myem}
\newcounter{mysubequation}[equation]

\makeatletter

\renewcommand{\section}{\@startsection{section}{1}{0em}{-\baselineskip}%
{\baselineskip}{\normalfont\large\bfseries}}
\renewcommand{\subsection}%
{\@startsection{subsection}{2}{0em}{-0.7\baselineskip}%
{0.7\baselineskip}{\normalfont\bfseries}}
\makeatother

\newcommand{\bi}{\begin{itemize}}
\newcommand{\ei}{\end{itemize}}

\newcommand{\be}{\begin{equation}}
\newcommand{\ee}{\end{equation}}
\newcommand{\bea}{\begin{eqnarray}}
\newcommand{\eea}{\end{eqnarray}}

\newcommand{\ldm}{\Delta m_{31}^2}
\newcommand{\sdm}{\Delta m_{21}^2}
\newcommand{\deltacp}{\delta_{\mathrm{CP}}}
\newcommand{\stheta}{\sin^2  2 \theta_{13} }
\newcommand{\ie}{{\it i.e.}}

\newcommand{\eg}{{\it e.g.}}

\newcommand{\cf}{{\it cf.}}

\newcommand{\eq}{Eq.}

\newcommand{\fig}{Fig.}

\newcommand{\Ref}{Ref.}
\newcommand{\Refs}{Refs.}
\newcommand{\Sec}{Sec.}

\newcommand{\App}{Appendix}

\newcommand{\Tab}{Table}

\newcommand{\equ}[1]{\eq~(\ref{equ:#1})}
\newcommand{\figu}[1]{\fig~\ref{fig:#1}}
\begin{document}
\begin{titlepage}
\renewcommand{\thefootnote}{\alph{footnote}}
\vspace*{-3.cm}
\begin{flushright}
\end{flushright}
\renewcommand{\thefootnote}{\fnsymbol{footnote}}
\setcounter{footnote}{-1}
{\begin{center}
{\large\bf Physics with a very long neutrino factory baseline
} \end{center}}

\renewcommand{\thefootnote}{\alph{footnote}}

\vspace*{.8cm}


{\begin{center} {\large{\sc
                Raj~Gandhi\footnote[1]{\makebox[1.cm]{Email:}
                raj@mri.ernet.in},~
                Walter~Winter\footnote[2]{\makebox[1.cm]{Email:}
                winter@physik.uni-wuerzburg.de}
                }}

\end{center}}

\vspace*{0cm}

{\it
\begin{center}

\footnotemark[1]
       Harish-Chandra Research Institute, \\
       Jhusi, Allahabad--211 019, India 

\footnotemark[2]%
       Institut f{\"u}r Theoretische Physik und Astrophysik, Universit{\"a}t W{\"u}rzburg \\
       D-97074 W{\"u}rzburg, Germany

\vspace*{1cm}

\today

\end{center}}

\vspace*{1.5cm}

{\Large \bf

\begin{center} Abstract \end{center}  }

We discuss the neutrino oscillation physics of a very long neutrino factory baseline over a broad range 
of lengths (between $6 \, 000$~km and 
$9 \, 000$~km), centered on the  ``magic baseline'' ($\sim 7 \, 500$~km) where correlations with the leptonic CP phase are suppressed by matter effects. Since the magic baseline depends only on the density, we study the impact of matter density profile effects and  density uncertainties over this range, and the impact of  detector locations off the optimal baseline. We find  that the optimal constant density describing the physics over this entire baseline range is about 5\% higher than the average matter density. This implies that the magic baseline is significantly shorter than previously inferred.
 However, while a single detector  optimization requires fine-tuning of the (very long) baseline length, its combination with a near  detector at a shorter baseline is much less sensitive to the far detector  location and to uncertainties in the matter density.
 In addition, we point out different applications of this baseline which go beyond its excellent correlation and degeneracy resolution potential. 
  We demonstrate that such a long baseline assists in the improvement of 
the $\theta_{13}$ precision  and in the resolution of  the octant degeneracy. Moreover,
we show that the neutrino data from such a baseline could be used to extract the matter density along the profile up to 0.24\% at $1 \sigma$ for large
$\stheta$, providing a useful discriminator between  different geophysical models.

\vspace*{.5cm}

\end{titlepage}

\newpage

\renewcommand{\thefootnote}{\arabic{footnote}}
\setcounter{footnote}{0}



\section{Introduction}

 Input from experiments is central to the search for a theory beyond the Standard Model of elementary particle physics. Given the extraordinary robustness of this
model, such input necessarily requires precision measurements of the relevant
parameters in order to reveal discrepancies, which, in turn, would provide clues 
to a higher theory. It is evident that at the  present time, in addition to
the upcoming Large Hadron Collider~\cite{Ellis:2006hd}, neutrino physics 
provides an unprecedented opportunity to move successfully towards
this goal via increasingly accurate  measurements of neutrino mass squared differences
 and mixing parameters. These parameters have only
been partially revealed by experiments so far (see, \eg,  \Refs~\cite{Schwetz:2006dh,Goswami:2006hc,Fogli:2006qg,Maltoni:2004ei}).  
A large number of experiments are thus planned or currently
under way to achieve this goal. Among these, in the near future, are neutrino
beam experiments~\cite{Ables:1995wq,Itow:2001ee,Ayres:2004js}   and
reactor experiments~\cite{Minakata:2002jv,Huber:2003pm,Anderson:2004pk,Ardellier:2004ui}. 
What may be learnt from these experiments depends, however,  on how large the mixing angle $\theta_{13}$
is. In particular, some of the important issues crucial to the formulation of a unified theory, such as the value of the CP violating parameter $\deltacp$, or the nature of the neutrino mass
hierarchy, may not be determined by these experiments.  With present bounds ($\sin^2\theta_{13}=0^{+0.047}_{-0} \, , \quad \deltacp=0^{+\pi}_{-\pi}$~\cite{Maltoni:2004ei}),  already restricting $\theta_{13}$ fairly stringently, it could,
in principle, be very small. This would inevitably result in the
  determination of  important  physics being relegated to an advanced future facility, such as a
  neutrino factory~\cite{Geer:1998iz,Albright:2000xi,Apollonio:2002en}.

The unprecedented reach and accuracy of a neutrino factory, and the 
broad scope of physics that can be explored  has been discussed in detail in 
several studies (see, \eg, \Refs~\cite{Barger:1999jj,Barger:1999fs,Barger:2000cp,Cervera:2000kp,Burguet-Castell:2001ez,%
Burguet-Castell:2002qx,Huber:2002mx,Huber:2006wb}). Much of the science
centers around isolating the
matter  effects in neutrino oscillations and determining how  the various degeneracies and correlations in the measured quantities can be successfully resolved to 
obtain unambiguous physics. The presence of intertwined parameters and effects
thus requires careful phenomenological and optimization studies in order to best determine the
specifications of the source facility and the location of detectors.  
 One of the significant results of these efforts is
the identification of the ``magic baseline'' for neutrino passage through the
Earth, which is a baseline between about $7 \, 000$ and $7 \, 500 \, \mathrm{km}$~\cite{Barger:2001yr,Huber:2003ak}. Clean measurements of $\stheta$ 
and the mass hierarchy
are possible due to the disappearance of CP effects at this baseline.
Most recently, these issues have been studied in detail from the point of view 
of the optimization with respect to  $\stheta$, the mass hierarchy, and CP
violation sensitivities, as well as the  measurement of the leading
atmospheric parameters in \Ref~\cite{Huber:2006wb}.  The conclusion from this study 
is that an optimal advanced neutrino factory would be operated with two
detector locations, one in the range between $3 \, 000 \, \mathrm{km}$ to 
$5 \, 000 \, \mathrm{km}$, and the second 
at or around the magic baseline. 
We recall that the magic baseline depends
only on the matter density, arising from the condition $\sqrt{2}
G_F n_e L = 2 \pi$, where $G_F$ is the weak coupling constant and $n_e$ is the
electron number density. Obviously, the exact density profile plays an important role 
for the physics potential of the very long baseline, whereas the oscillation parameter values and 
neutrino energies are (almost) irrelevant. In comparison, the optimal ``short'' baseline 
length depends on the oscillation parameters, but the baseline window is large enough
such that one does not have to worry too much about the actual parameters and the
matter density profile for the choice of the detector location.

Since the precise location of a detector is a function of many variables,
a good fraction of which may be unrelated to physics, it is necessary to 
study the impact of locating the second detector not at the optimal magic
length, but somewhere in the range between $6 \, 000 \, \mathrm{km}$ and 
$9 \, 000 \, \mathrm{km}$. In addition to this 
geographical uncertainty, the effects of the matter density profile and its
uncertainties on the optimization are significant. In this study, we explore these 
questions by introducing a realistic matter density profile model in \Sec~\ref{sec:model}, 
and by examining the physics consequences of  this model in \Sec~\ref{sec:profile}. 
Note that in our scenario, the neutrino factory feeds 
two detectors, one at $L_1=4 \, 000 \, \mathrm{km}$~\cite{Huber:2006wb},
the other at $L_2$ between $6 \, 000 \, \mathrm{km}$ and $9 \, 000 \, \mathrm{km}$.
We study, as new applications, the $\theta_{13}$ precision measurement in 
\Sec~\ref{sec:theta13} and the 
octant degeneracy resolution in \Sec~\ref{sec:resdeg}, and we sketch the physics 
case for such a baseline in \Sec~\ref{sec:physics}. 
We believe that the investigation of these questions is especially germane in
the light of the presently ongoing International Scoping Study for a future
neutrino factory and superbeam facility~\cite{ISSdetectorWG}. As an
interesting application for geophysics, we also demonstrate how neutrino
measurements may help to extract information about the matter density 
profile and to discriminate between different Earth density models in 
\Sec~\ref{sec:matterdensity}.  
 
\section{Physics at the Magic Baseline}
\label{sec:pheno}

It was noticed in \Ref~\cite{Barger:2001yr,Huber:2003ak} that the condition $\sqrt{2}
G_F n_e L = 2 \pi$ leads to a disappearance of CP violation
effects and related degeneracies in $P_{e\mu}$.  Several studies have
subsequently explored the physics possibilities resulting from the enhanced matter effects
at and around this baseline, see, \eg, \Refs~\cite{Gandhi:2004bj,Gandhi:2004md,Akhmedov:2004ny,Agarwalla:2005we,Donini:2005db,Huber:2006wb,Agarwalla:2006vf}. A recent 
closer examination of the phenomenology has been done in \Ref~\cite{Smirnov:2006sm}.
In this section we review the salient features and reasons which make this
baseline a phenomenologically attractive one. We start with an approximate 
analytical expression for $P_{e\mu}$, which is accurate up to second order in
the combination of two parameters which can  (usually) be treated as small,
namely $\alpha\equiv\frac{\Delta m^2_{21}}{\Delta m^2_{31}}\sim \pm 0.03$
and $\sin 2\theta_{13}$~\cite{Cervera:2000kp,Freund:2001pn,Akhmedov:2004ny}:  
\begin{eqnarray}
P_{e\mu} & \simeq & \sin^2 2\theta_{13} \, \sin^2 \theta_{23} \frac{\sin^2[(1- \hat{A}){\Delta}]}{(1-\hat{A})^2} 
\nonumber \\
&\pm&   \alpha  \sin 2\theta_{13} \,  \sin 2\theta_{12} \, \sin 2\theta_{23} \, \sin \delta_{\mathrm{CP}}   \, 
\sin({\Delta})  \frac{\sin(\hat{A}{\Delta})}{\hat{A}}  \frac{\sin[(1-\hat{A}){\Delta}]}{(1-\hat{A})}
\nonumber  \\
&+&   \alpha  \sin 2\theta_{13} \,  \sin 2\theta_{12} \, \sin 2\theta_{23} \, \cos \delta_{\mathrm{CP}} \,  \cos({\Delta})  \frac{\sin(\hat{A}{\Delta})}{\hat{A}}  \frac{\sin[(1-\hat{A}){\Delta}]} {(1-\hat{A})}
 \nonumber  \\
&+&  \alpha^2 \, \cos^2 \theta_{23}  \sin^2 2\theta_{12} \frac{\sin^2(\hat{A}{\Delta})}{\hat{A}^2}.
\label{equ:PROBMATTER}
\end{eqnarray}
In the above, $\Delta\equiv\frac{\Delta m^2_{31} \,  L}{4 \, E_{\nu}}$ and
$\hat{A}\equiv\frac{2 \sqrt{2} G_F n_e E}{\Delta m^2_{31}}$ and the $\theta_{ij}
$ are the usual neutrino mixings. The defining condition for the magic
baseline is 
\begin{equation}
    \sin ( \hat{A} \Delta ) = 0 \quad \Rightarrow \quad \frac{G_F n_e L}{\sqrt{2}} =  \pi 
\label{equ:magic}
\end{equation}
for the first (shortest baseline) non-trivial solution.
 It is apparent that in this case only the first term
in \equ{PROBMATTER} remains, which means that any dependence on $\deltacp$ is
removed. It is important to  note that \equ{magic} is also the defining equation for the
    refraction length of a medium~\cite{Wolfenstein:1977ue}. For large
    matter  densities and high energies beyond the resonance energy of the
    medium, the effective neutrino oscillation length in matter approaches the 
refraction length. Moreover, at these baselines and energies, the oscillation length in
    matter is driven by $\Delta m^2_{21}$, and not $\Delta m^2_{31}$.    
When the matter oscillation length equals the refraction length, the 
     phase in matter for oscillations driven by the solar mass
    difference becomes $2\pi$, and terms containing  $\Delta m^2_{21}$
and $\deltacp$ in the above expansion drop out of the probability~\cite{Smirnov:2006sm}.

 We  stress  that these  statements related to the magic baseline are  subject to corrections from the following: a) The accuracy and the validity of the above expansion in $\alpha$ and $\sin 2\theta_{13}$, b) the validity of the constant density approximation, c) our knowledge of the Earth's 
density profile, and d) the correctness of choosing the average density  as the constant density which determines the magic baseline via \equ{magic} above.

The accuracy and validity of various analytical series expansions for matter
probabilities has been discussed in 
\Ref~\cite{Akhmedov:2004ny}. In general, one can  choose to
expand either in $\alpha$ alone, or in $\sin 2\theta_{13}$ alone, retaining terms 
up to first order in one of these parameters while treating the other parameter exactly.
Alternatively, one can choose to treat both as small parameters (\ie, do a double expansion),
and keep terms up to second order, as we have done  
above in \equ{PROBMATTER}. The single expansion in $\alpha$ retains
the exact dependence in $\sin 2\theta_{13}$, and hence is more accurate than the
double expansion for values of $\sin 2 \theta_{13}$ close to the upper bound. 
It is valid for $\alpha \, \Delta\equiv\frac{\Delta m^2_{21} \, L}{4 \, E_{\nu}} \ll 1$
(which translates to $L/E \ll 10^4$ km/GeV), \ie, 
when the vacuum oscillation length defined by   the ``solar'' mass
squared difference is much larger than $L$. On the other
hand, the single expansion in $\sin 2\theta_{13}$ is most accurate when this
parameter assumes values which are very small ($\sim 10^{-3}$ or less). The
double expansion in \equ{PROBMATTER} is, in general, robust over a wide range of parameters
(in particular, $\sin 2\theta_{13}$, $L$, and $E$)
in the sense that relative errors in $P_{\mu e}$, $P_{\mu\tau}$, and $P_{\mu\mu}$
are restricted to be below $5\%$. In terms of validity, 
errors resulting from its use are lowest (below $1\%$) when  $\alpha \, \Delta \equiv
\frac{\Delta m^2_{21}\, L}{4 \, 
  E_{\nu}} \ll 1$, \ie,  $L/E  \ll 10^4$ km/GeV, and when $\sin 2\theta_{13}$ is small but 
still above the range which makes the single expansion in it preferable.
As far as the defining equation for the magic baseline
\equ{magic} is concerned,  note that it is directly obtained from the double expansion. Therefore,
 these constraints on the 
validity of the expansion need to be kept in mind when drawing conclusions about results 
from a detector located at $L=L_{\mathrm{Magic}}$, especially when precision of a few percent is
important.\footnote{An approach independent of an  analytical 
expansion would numerically search for the minimal dependence on the solar amplitude
contributions to the transition probability, but it would still be subject to
uncertainties in our knowledge of the density profile.} 

As a first approximation, \equ{magic} can be usefully rewritten as
\begin{equation}
    L_{\mathrm{Magic}} \simeq \frac{\sqrt{2}\pi}{ G_F n_e} \simeq 7 \, 800 \, \frac{4.2 \, \mathrm{g/cm^3}}{\bar{\rho}(L)} \, \mathrm{km}
\label{equ:magic2}
\end{equation}
where 
\begin{equation}
 \bar{\rho}(L) = \frac{1}{L} \int\limits_{0}^{L} \rho(x) dx
 \label{equ:rhoavg}
\end{equation}
is the baseline-averaged density. Solving this equation, we find
 $L_{\mathrm{Magic}} \simeq 7 \, 630 \, \mathrm{km}$ and 
$\bar{\rho}( 7 \, 630 \, \mathrm{km}) \simeq 4.29 \, \mathrm{g \, cm^{-3}}$\footnote{ Note the non-trivial inter-dependance inherent in seeking a solution to  \equ{magic2}, since 
 $\bar{\rho}(L)$ in the equation  depends actually on $L_{\mathrm{Magic}}$.}
  Most studies in the literature have used
$\bar{\rho}(L)$ in calculations, and have drawn conclusions based on it as the preferred choice for  the constant density. 
There is no doubt that the simplified analysis  based
on the constant density approximation has provided many insights into the effect of matter
oscillations on neutrinos as they travel through the Earth. 
However, as precision assumes increasing importance in the
determination of neutrino parameters, it is important to ask how well    
$\bar{\rho}(L)$ reproduces the profile effect for a given baseline. 
In particular, we will discuss if $\bar{\rho}(L)$ is indeed the best choice
for the constant density to reproduce physics. We will introduce a 
constant reference density for that purpose which can be different from $\bar{\rho}(L)$,
and examine its dependence on the baseline, on the neutrino oscillation parameters 
(such as $\stheta$ and $\deltacp$), and on the oscillation channel that one is observing.
In the next section, we describe the simulation methods employed by us, prior
to addressing this question in detail in \Sec~\ref{sec:model}, where we also describe our
procedure for modeling the Earth's density profile.
 
Finally, we stress two important features of the magic baseline relevant  to  measurements of the Earth's matter density. First, the clean dependence of physics at the magic baseline on the matter density due
to the independence of CP and solar mass-squared difference terms suggests that
it could be an optimal baseline for Earth density measurements. In other
words, if the density was considered as just another parameter to be determined
via neutrino experiments, this determination would be cleanest and most
uncluttered by correlations if it was done at the magic baseline. Secondly, we note that the condition 
$\sin ( \hat{A} \Delta ) = 0$ automatically  tells us that the sensitivity to any change in the assumed constant  density (\ie, the derivative) is maximal here for the terms proportional to $ \sin (\hat{A} \Delta)$, and consequently, is expected to be high for the full probability. This fact  can be usefully employed  to extract information about the density.  These ideas are 
explored further in \Sec~\ref{sec:matterdensity} of this work. 

\section{Simulation methods}
\label{sec:methods}

Our simulation is based upon the standard neutrino factory in \Refs~\cite{Huber:2002mx,Huber:2006wb}
with a muon energy $E_\mu = 50 \, \mathrm{GeV}$ and a $50 \, \mathrm{kt}$ magnetized iron detector. We use a total luminosity of $4.24 \times 10^{21}$ useful muon decays and $4.24 \times 10^{21}$ useful anti-muon decays, which can be achieved by four years of operation in each polarity at $1.06 \times 10^{21}$ useful parent decays/year, or by eight years of simultaneous operation of both polarities at $0.53 \times 10^{21}$ useful parent decays/year/polarity. The neutrino factory uses both the muon neutrino/anti-neutrino appearance ($\nu_e \rightarrow \nu_\mu$) and disappearance ($\nu_\mu \rightarrow \nu_\mu$) channels. As in \Ref~\cite{Huber:2006wb}, we use a data sample without charge identification for the disappearance channels to avoid cuts from the charge identification, and therefore increase statistics. For a more detailed description, see \Refs~\cite{Huber:2002mx,Huber:2006wb}.

In order to perform the simulation, we use the GLoBES software~\cite{Huber:2004ka}. Since it is relevant for this study, let us focus on the way GLoBES treats the matter density. In GLoBES, the matter density of each experiment is implemented as yet another oscillation parameter, \ie, the total systematics $\chi^2$ for one experiment is a function of the oscillation parameters and the matter density. For two experiments, the
systematics $\chi^2_{\mathrm{sys}}$ is then computed as
\begin{eqnarray}
\chi^2_{\mathrm{sys}}(\theta_{12},\theta_{13},\theta_{23},\deltacp,\sdm,\ldm,\hat{\rho}_1, \hat{\rho}_2) & = &
 \chi^2_{L_1}(\theta_{12},\theta_{13},\theta_{23},\deltacp,\sdm,\ldm,\hat{\rho}_1) + \nonumber \\ & + & \chi^2_{L_2}(\theta_{12},\theta_{13},\theta_{23},\deltacp,\sdm,\ldm,\hat{\rho}_2)
\end{eqnarray}
for two different baselines $L_1$ and $L_2$ with different density scaling factors $\hat{\rho}_1$ and $\hat{\rho}_2$, where the oscillation probabilities in $\chi^2_{L_1}$ and $\chi^2_{L_2}$ depend on the respective density scaling factors. GLoBES allows arbitrary density profiles, which are scaled by these density scaling factors $\hat{\rho}$. For instance, for a constant density profile with density $\rho_0$, the actual density will be computed as $\hat{\rho} \cdot \rho_0$. For an arbitrary profile, $\hat{\rho}$ acts as an overall normalization factor with which the density in each layer is multiplied.  As the
next step in the $\chi^2$ calculation, the external input is added as
\begin{equation}
\tilde{\chi}^2 = \chi^2_{\mathrm{sys}}(\theta_{12},\theta_{13},\theta_{23},\deltacp,\sdm,\ldm,\hat{\rho}_1,\hat{\rho}_2)
+ \left( \frac{\hat{\rho}_1 - 1.0}{\sigma_{\hat{\rho}_1}} \right)^2 +  \left( \frac{\hat{\rho}_2 - 1.0}{\sigma_{\hat{\rho}_2}} \right)^2 + \hdots \, , 
\end{equation}
where the dots correspond to other potential external constraints (such as for the solar parameters).
Typically, we use $\sigma_{\hat{\rho}} = 0.05$, corresponding to a 5\% overall normalization uncertainty of the 
density profile. This $\tilde{\chi}^2$ is then marginalized over the unwanted oscillation parameters and density scaling factors.
The marginalization procedure corresponds to the projection of the eight-dimensional fit manifold
onto the targeted sub-space. For example, for a $\stheta$ precision measurement, all other parameters and density scaling factors are marginalized over. For a $\hat{\rho}_2$ precision measurement (corresponding to the relative error for a constant density), all oscillation parameters {\em and} $\hat{\rho}_1$ are marginalized over, as well as $\sigma_{\hat{\rho}_2} \rightarrow \infty$.\footnote{Note that in this case, we still impose an uncertainty on $\hat{\rho}_1$, because $\hat{\rho}_1$ is, in the most conservative case, fully uncorrelated with $\hat{\rho}_2$.}
Therefore, there is no {\em a priori} difference between the oscillation parameters and density scaling
 factors. 

In some cases, we will even use a more detailed approach with different parts of the density
profile being treated with different scaling factors. In this case, one has to perform the density 
marginalizations in GLoBES manually, \ie, set the density profile in each step and manually scan the density
 scaling factors
of the different sub-profiles. For example, we will use this approach to measure the lower mantle density only, while we impose an uncertainty on the upper mantle density of the Earth. 

For the oscillation parameters, we use $\sin^2 2 \theta_{12}=0.83$, $\sin^2 2 \theta_{23} = 1$, $\sdm = 8.2 \cdot 10^{-5} \, \mathrm{eV}^2$, $\ldm = 2.5 \cdot 10^{-3} \, \mathrm{eV}^2$, and a normal hierarchy unless stated otherwise (see, \eg, \Refs~\cite{Maltoni:2004ei,Fogli:2003th,Bahcall:2004ut,Bandyopadhyay:2004da,Schwetz:2006dh}).
In addition, we assume a 5\% external measurement for $\sdm$ and
$\theta_{12}$ from solar experiments (see, \eg,
\Ref~\cite{Bahcall:2004ut}). We furthermore include matter density uncertainties
of the order of 5\%~\cite{Geller:2001ix,Ohlsson:2003ip} uncorrelated
between the different baselines -- unless we measure the matter density.
In principle, we include all parameter correlations and discrete degeneracies~\cite{Barger:2001yr,Burguet-Castell:2001ez,Minakata:2001qm,Fogli:1996pv}
where applicable. However, the octant degeneracy will not be present for maximal mixing.

\section{Modeling the density profile and determining the optimal constant density}

\label{sec:model}

 Most  simulations in the literature   so far have used  the baseline-averaged
 matter density calculated using \equ{rhoavg}, 
 where $\rho(x)$ is calculated   along the baseline using 
 the PREM profile. Since the magic baseline is mainly determined
by the matter density, this description may not be accurate enough to find
the optimal detector location (see, \eg, \Ref~\cite{Ohlsson:2001et} for a
discussion at the probability level). Therefore, we follow a different
approach in this section, and seek a constant density that best simulates the
PREM profile. We label this density as the ``reference density'', $\rho_{\mathrm{Ref}}$.

\begin{figure}[t]
\begin{center}
\includegraphics[width=\textwidth]{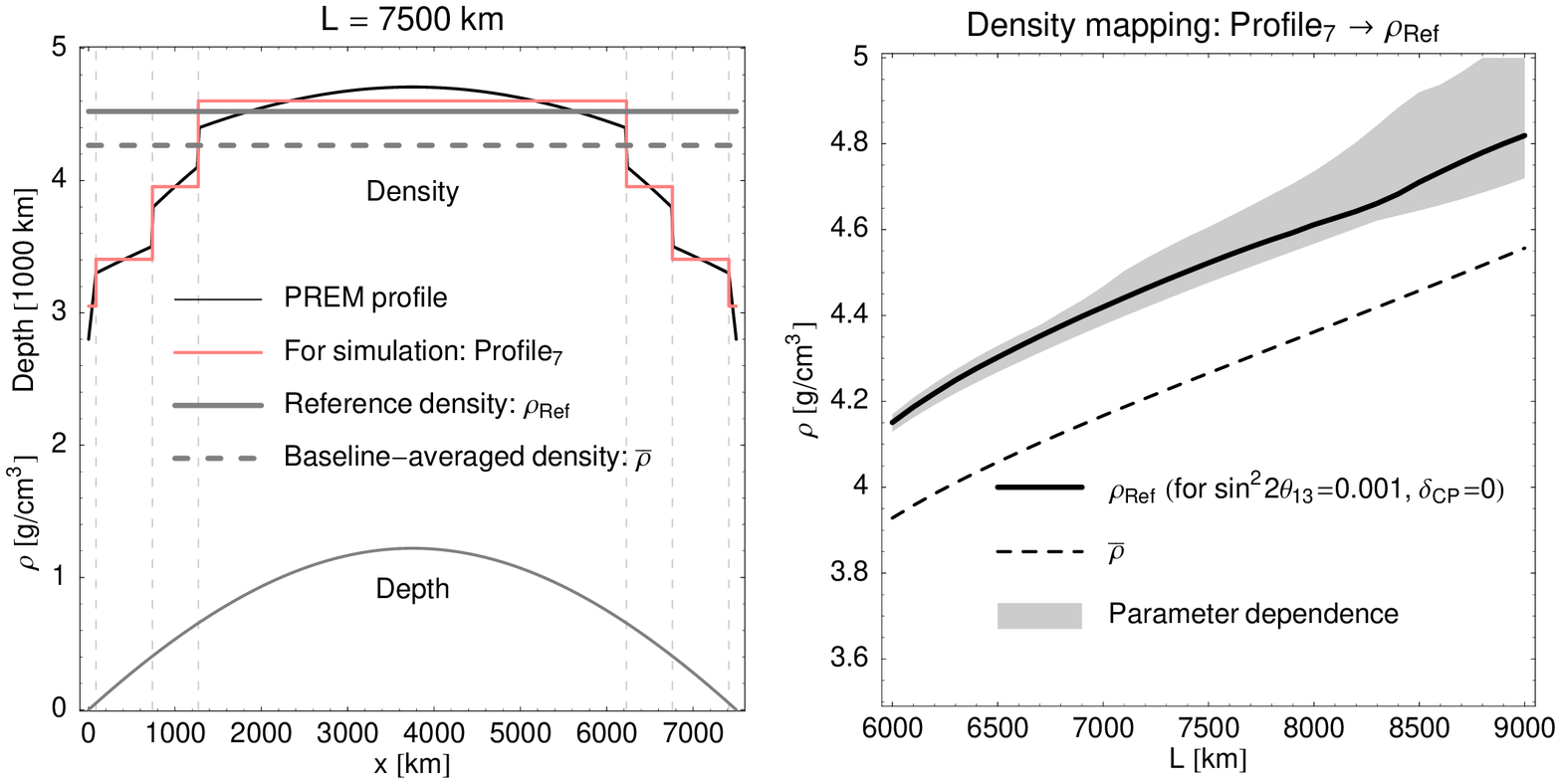}
\end{center}
\mycaption{\label{fig:densmapping} Mapping from PREM profile via the simulation profile ``Profile$_7$'' to our constant reference density $\rho_{\mathrm{Ref}}$. The left plot shows several possible matter density profiles as function of the propagation length $x$ for a fixed baseline $L=7 \, 500 \, \mathrm{km}$, as well as the depth as function of baseline. The right plot shows our constant reference density $\rho_{\mathrm{Ref}}$ corresponding to Profile$_7$ in the left plot as function of the baseline $L$ (solid curve) -- see text for explanation of the mapping. The baseline-averaged density $\bar{\rho}$ is shown for comparison as well, and the parameter dependence of the mapping on $\stheta$ and $\deltacp$ is illustrated by the gray-shaded area.
}
\end{figure}

To begin with, we note that a numerical simulation cannot use the unmodified
PREM profile because most numerical techniques are based on the
re-diagonalization of the Hamiltonian in constant matter density (\cf, for
instance, \Ref~\cite{Ohlsson:1999um}). Therefore, the PREM profile has to be
made accessible to numerical simulations, \ie, it has to be accurately
modeled. One possibility  is to use a Fourier expansion~\cite{Ota:2000hf}, which, however, does not describe the edges very accurately
and is not easily implementable in GLoBES. Therefore, we use a different approach,
which is suitable for the baseline window of interest here ($6 \, 000 \,
\mathrm{km}$ to $9 \, 000 \, \mathrm{km}$). We illustrate this model, which we
call``Profile$_7$'' (because of seven layers) in \figu{densmapping}  and
show how it compares  to the PREM profile. For any baseline between $6 \, 000
\, \mathrm{km}$ and $9 \, 000 \, \mathrm{km}$, the profile looks very similar
to the one for $L=7500$ km shown in the left panel of
\figu{densmapping}.  We compute Profile$_7$ for any given $L$ in
this range by choosing the density jumps as edges between two layers (such as between
upper mantle and lower mantle of the Earth). Each
layer is then simulated  by using the average density given by \equ{rhoavg}
within that layer. Note that although the use of more steps would make the
modeling more accurate,  any realistic simulation seeks to use an optimal number that
retains accuracy while keeping the computational time within reasonable limits.
We have checked that Profile$_7$ reproduces the results of a simulation with significantly more steps
with sufficient accuracy.

\begin{figure}[t]
\begin{center}
\includegraphics[width=\textwidth]{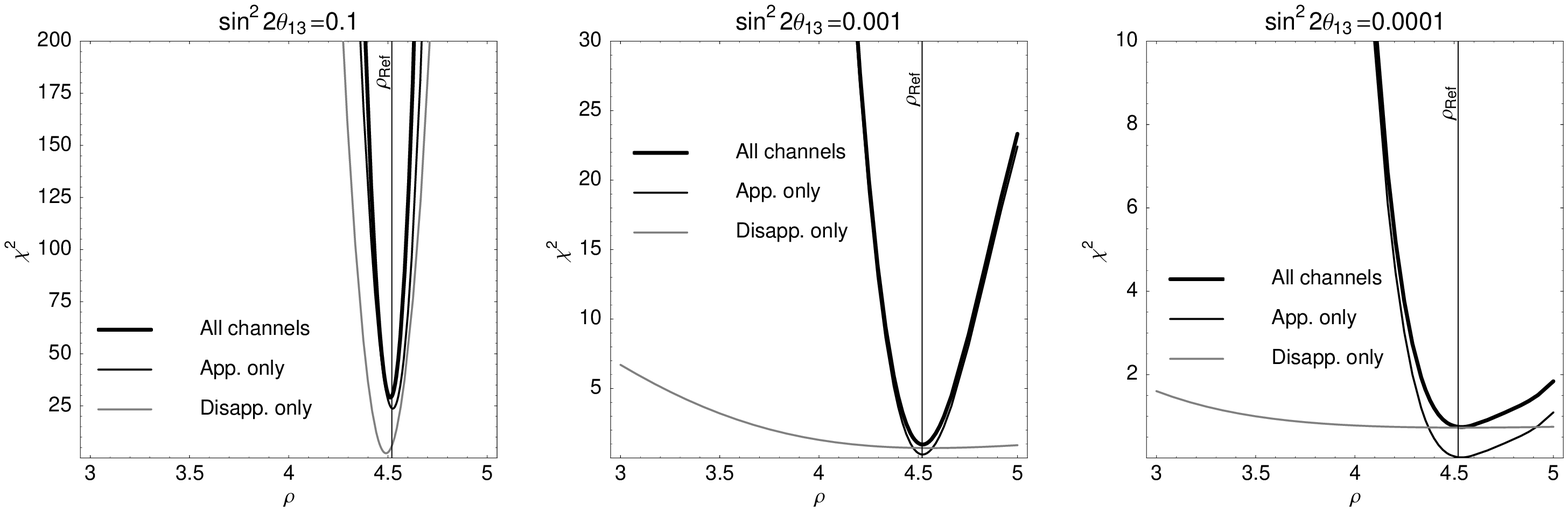}
\end{center}
\mycaption{\label{fig:chandep} $\chi^2$ between a simulated Profile$_7$ and a fit constant
density profile as a function of the fit density $\rho$. The different panels
correspond to three different values of $\stheta$ (see plot labels). In each plot, we show 
three different oscillation
  channel choices (appearance only, disappearance only and all channels combined). The 
reference density sits at the minimum of the $\chi^2$ for all channels
  combined (chosen for $\stheta=0.001$), and is marked by the vertical lines in all panels. 
Note that the scales for the vertical axes differ from each other
  in the three panels. In this figure, $\deltacp=0$ and $L=7 \, 500 \, \mathrm{km}$.}
\end{figure}

As mentioned above, we seek an optimal constant density which can be
reliably used as a reference density in lieu of full numerical simulations. We define this to be the
 constant density which simulates the PREM profile best
and denote it as  $\rho_{\mathrm{Ref}}$. In order to relate
$\rho_{\mathrm{Ref}}$ to Profile$_7$ for any given baseline, we use the
$\chi^2$ from the complete neutrino factory simulation described in the
last section (including neutrino and antineutrino appearance and
disappearance). We identify  the $\rho_{\mathrm{Ref}}$ which  simulates Profile$_7$
best by requiring that  $\chi^2$ be  minimal between a constant fit profile with
$\rho_{\mathrm{Ref}}$ and the simulated Profile$_7$, for a given set of fixed oscillation
parameters. 
This means that we minimize the contribution from the leading (zeroth) order 
profile effect to the total $\chi^2$.
We illustrate this process in \figu{chandep}, in which we show
the variation of $\chi^2$ with the choice of the constant density $\rho$ for different 
values of $\stheta$ and for different oscillation channel choices. The reference density
is marked by the vertical lines as the minimum of the $\chi^2$ between the constant density
profile and Profile$_7$ (obtained for $\stheta=0.001$, $\deltacp=0$. In all three panels 
(each representing a different value of $\stheta$ ),
the appearance channel  dominates the determination of $\rho_{\mathrm{Ref}}$, but for 
large $\stheta$, the disappearance channel also contributes significantly.
For small $\stheta$, the appearance
 channel $\chi^2$ becomes much more sensitive to deviations from $\rho_{\mathrm{Ref}}$ than
the disappearance channel. The reason is that the disappearance channel is almost in the
two-flavor vacuum regime described by
$\nu_\mu \leftrightarrow \nu_\tau$ oscillations for $\stheta \rightarrow 0$, 
\ie, independent of matter effects. Therefore, inaccuracies in the estimation of the 
constant reference density are much less important for the disappearance channel than for
the appearance channel.

We show our choice for $\rho_{\mathrm{Ref}}(L)$ as function of baseline as the solid curve in
\figu{densmapping} (right), which was obtained for the reference values
$\stheta=0.001$ and $\deltacp=0$. 
Expectedly, this mapping  depends on the choice of oscillation parameters, as
 illustrated in \figu{densmapping} (right panel) by the shaded area obtained
for different sets of parameters. 
 We find that the  $\deltacp$-dependence
for large values of $\stheta$ is very mild, reflecting the dominance of matter
effects. As shown above,  the minimum $\chi^2$
between the profiles $\rho_{\mathrm{Ref}}$ and Profile$_7$ is rather large
($\chi^2 \sim 15$ to $35$) and the function $\chi^2(\rho)$ is rather steep for the large $\stheta$ case. On the other hand, for small values of $\stheta$, the dependence on $\deltacp$ is stronger, but the minimum $\chi^2$ between the profiles $\rho_{\mathrm{Ref}}$ and Profile$_7$ is rather small ($\chi^2 \sim 1$ to $10$), while the function $\chi^2(\rho)$ is more shallow. Therefore, we conclude that the parameter dependence in \figu{densmapping}, right (shaded area) may actually not be very significant,
either because there is no significant $\deltacp$-dependence (large $\stheta$), or because the
statistical contribution from deviations is small (small $\stheta$). 
We will test this hypothesis later in full numerical simulations using different profiles. Note that we will simulate any baseline shorter than $6 \, 000 \, \mathrm{km}$ as usual with $\bar{\rho}$.

A noteworthy point from \figu{densmapping} is that
$\rho_{\mathrm{Ref}}$ 
is consistently higher than $\bar{\rho}(L)$ by about $5\%$. This
reflects the dominance of high density regions in the lower mantle (\cf, left panel of 
\figu{densmapping}). The fact
that matter effects at short baselines are relatively suppressed (irrespective
of the density profile) has been discussed in \Ref~\cite{Akhmedov:2000cs}.
Thus, in our case, the high density central region of the profile contributes with a higher
weight compared to the low density regions. At short baselines, transition
amplitudes for neutrinos can be treated perturbatively, and the leading order
terms reflect the contributions of the matter independent off-diagonal terms of the 
effective Hamiltonian, whereas matter contributions 
reside in the diagonal terms. None of this information finds its way into the
simple averaging that determines $\bar{\rho}(L)$. Clearly, an optimized constant density for 
each channel offers superior accuracy compared to simple averaging since it 
incorporates the effects of matter on transition amplitudes in a
weighted way. This has significant  consequences for determining the
length of the magic baseline. For instance, we see from
\figu{densmapping}
that the $\rho_{\mathrm{Ref}}$ corresponding to $7 \, 600$~km is $4.5 \, \mathrm{g/cm^3}$ rather than 
$4.2 \, \mathrm{g/cm^3}$. Therefore, the magic baseline length becomes an estimated $\sim 7 \, 300$ km 
according to \equ{magic2}, \ie, several hundred kilometers shorter. It is thus important that prior 
to fixing the location of the far detector, an optimized reference density be
  used as opposed to using $\bar{\rho}(L)$. This sensitivity to the
  density changes if two detectors are used, one at $L_1 = 4 \, 000 \, \mathrm{km}$, 
and the other in the range between $6 \, 000$~km and $9 \, 000$~km. We
  study this combination in the next section.  

\section{How precisely does one have to tune very long baseline length?}
\label{sec:profile}

In \Ref~\cite{Huber:2006wb}, the combination of  a short baseline $L \simeq 4 \, 000 \, \mathrm{km}$ optimal for CP violation and the ``magic baseline'' $L \simeq 7 \, 500 \, \mathrm{km}$ optimal for degeneracy resolution and mass hierarchy determination was found to have an excellent
physics potential. The choice of this specific $L=7 \, 500 \, \mathrm{km}$ was  (in this and earlier studies~\cite{Huber:2003ak}) based on the $\stheta$ sensitivity being optimal there due to vanishing
$\deltacp$ effects. While it was found that the short baseline is rather insensitive to the
specific baseline length in a window between about $3 \, 000$ and $5 \, 000 \, \mathrm{km}$, 
the magic baseline optimum was found in a very small window of $L$ which only
depends on the matter density profile. In this section, we therefore discuss
how precisely one has to place the detector on the magic baseline, a question
very relevant for the selection of detector locations, and the extent of the
impact  from profile effects and inaccurately estimated  profiles. 

Before we actually discuss the impact of the long baseline choice on the physics potential,
let us focus on the most important questions:
\begin{enumerate}
\item
{\em Detector location:} How much sensitivity does one lose if one moves the
detector slightly off the optimal length?
\item
{\em Unknown matter density:} The length of the magic baseline depends on the matter density. What happens if the matter density along the baseline has been misjudged within current geophysical uncertainties?
\item
{\em Profile effects:} How well does a constant density simulate the matter density profile?
Does that affect the baseline optimization? Is the actual sensitivity better or worse compared to using
a constant density?
\end{enumerate}

In order to answer question~1, we will show the sensitivities in a window between $6 \, 000$ and $9 \, 000 \, \mathrm{km}$ in very small steps. For the complete picture, see, \eg, \Ref~\cite{Huber:2006wb}. For question~2, we will show the results for $\rho_{\mathrm{Ref}}$ as well as the most conservative estimates $0.95 \, \rho_{\mathrm{Ref}}$ and $1.05 \, \rho_{\mathrm{Ref}}$ from current seismic wave reconstructions. And for question~3, we will compare the results for $\rho_{\mathrm{Ref}}$ with the ones from Profile$_7$. In addition, we will show the curves for $\bar{\rho}$ to estimate how far earlier studies have been off the more realistic calculations.

\begin{figure}[t!]
\begin{center}
\includegraphics[width=\textwidth]{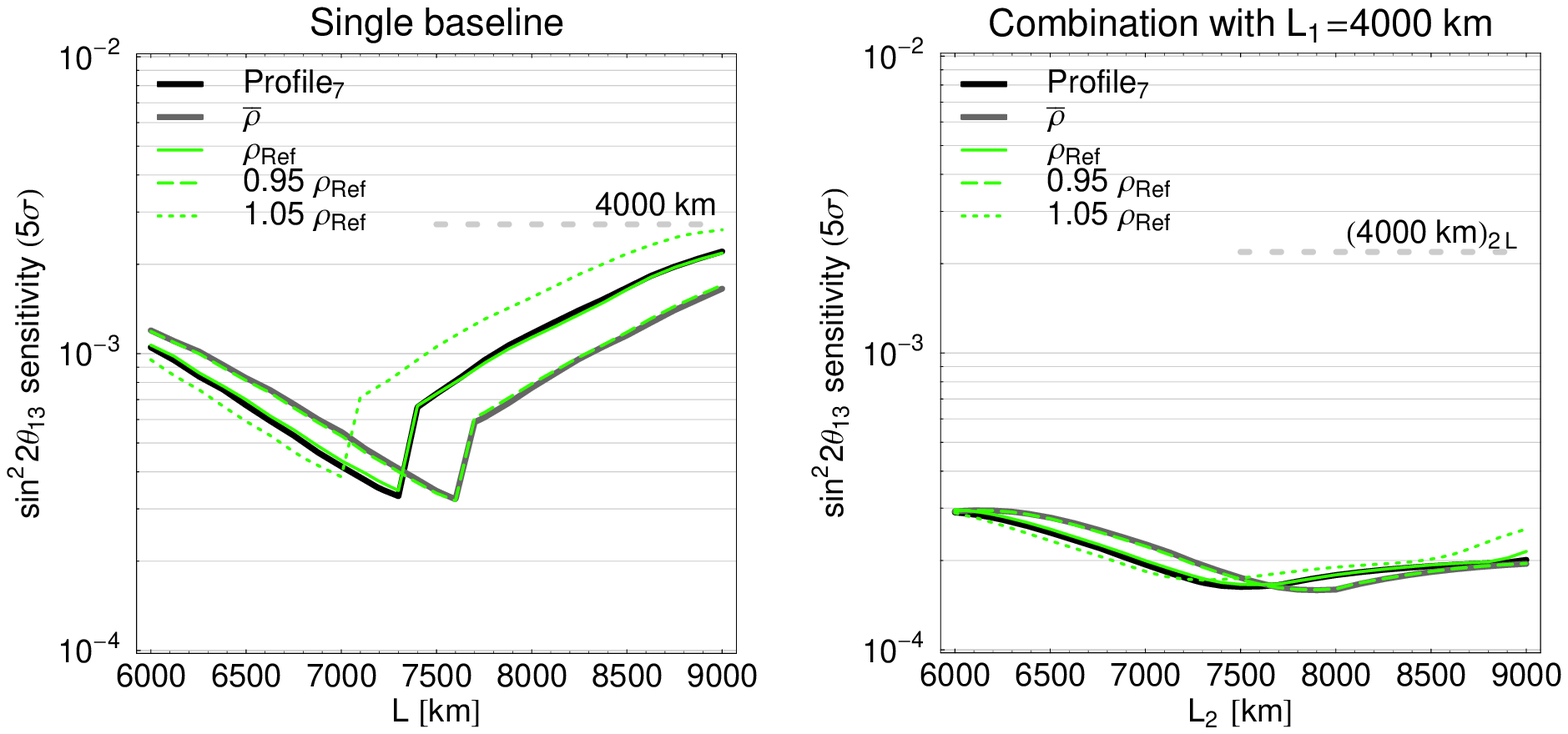}
\end{center}
\mycaption{\label{fig:theta13} The $\stheta$ sensitivity as function of $L$ including systematics, correlations, and degeneracies at the $5\sigma$ confidence level. The left plot corresponds to
a single baseline only, while the right plot shows the combination with a shorter baseline $L_1 = 4 \, 000 \, \mathrm{km}$. In both cases, placing all detector mass at $L = 4 \, 000 \, \mathrm{km}$ is shown for comparison as horizonal lines (the index ``2L'' refers to ``double luminosity'' because the detector mass in the right plot is twice as high as in the left plot). The different curves represent our different matter density models as represented in the plot legends.}
\end{figure}

We show in \figu{theta13} the $\stheta$ sensitivity as function of baseline. In the left panel, 
one can clearly see a minimum at about $7 \, 300 \, \mathrm{km}$ for Profile$_7$ from a single
baseline optimization. The reference density $\rho_{\mathrm{Ref}}$ simulates
the profile in an excellent way. Misjudging the matter density, however, can
affect the sensitivity quite severely, and a slightly shorter baseline of $L
\simeq 7 \, 000 \, \mathrm{km}$ may be safest from this point of
view.\footnote{For a shorter baseline $L \simeq 7 \, 000 \, \mathrm{km}$, the
  loss in sensitivity is much smaller when misjudging the matter density (such
  as actually having a 5\% higher density) than for a   baseline fine-tuned to $7 \, 300 \, \mathrm{km}$.} However, from \figu{theta13}, right panel, one can read off that
while the optimum is still at $L_2 \simeq 7\, 500 \, \mathrm{km}$, a baseline
$L_2 \simeq 7\, 500 ^{+1 \, 000 }_{-500 } \, \mathrm{km}$ does not affect the
combined result from two baselines significantly. Additionally,  the exact prediction for the matter density profile is irrelevant. The reason is that the very long baseline acts as a degeneracy resolver which can lift the intrinsic and mass hierarchy degeneracies far enough irrespective of the exact baseline choice. Note that placing all detector mass at $L_1 = 4 \, 000 \,
 \mathrm{km}$ is shown for comparison (horizontal lines), and one can clearly see the order of
 magnitude improvement coming from the second baseline. We have checked these results for the combination
 with $L_1 =3 \, 000 \, \mathrm{km}$, and also for  a larger value of $\ldm$. In both
 cases, none of the results change qualitatively. In addition, we show the impact of a smaller detector mass
 at $L_2$ in \App~\ref{app:detmass}. We conclude that for the $\stheta$ sensitivity, the detector location
 can be chosen quite freely if there is sufficient statistics from a shorter
 baseline.\footnote{This has implications for the proposed INO detector~\cite{Athar:2006yb}, the
   location of which will be at distances of $6560$ km from JHF and $7145$ km 
from CERN, thus making it a suitable very long  baseline detector.}

Assuming the $\stheta$ sensitivity as our primary criterion for optimization, one can cross-check the
results with the mass hierarchy and CP violation sensitivity. Note that an approach that assumes  mass hierarchy and CP violation as primary optimization criteria would be much more complicated because one can optimize for two different degrees of freedom: $\stheta$ reach and  $\deltacp$ values (``Fraction of $\deltacp$'' for which one can discover the mass hierarchy or CP violation).  In this study, we choose to use the fraction of $\deltacp$ for different physics scenarios in $\stheta$ to cross-check the optimization. However, unless  $\stheta \gtrsim 0.01$, it will not be possible to predict which scenario is actually realized in nature before a neutrino factory is built. Therefore, these physics scenarios can really only serve as  cross-checks to predict how well the final setup would perform for certain hypothetical choices of simulated values.

It has been demonstrated in \Ref~\cite{Huber:2006wb} that a very long baseline is absolutely necessary for the mass hierarchy determination if $\stheta$ is small. We have therefore checked the mass hierarchy sensitivity with this in mind and  have found that  hierarchy determination should be easily possible for  $L_1 = 4 \, 000 \, \mathrm{km}$ combined with $L_2$ in the window between $6 \, 000$ and $9 \, 000 \, \mathrm{km}$ for any value of $\deltacp$ and $\stheta \gtrsim 10^{-4}$ at the $3 \sigma$ confidence level (for any of the tested density models). Only at $\stheta \sim 10^{-4}$ does the baseline window $L_2 = (7 \, 300 \pm 200)  \, \mathrm{km}$ represent a local optimum (Fraction of $\deltacp$ = 100\%), but the loss in the fraction of $\deltacp$ if one moves away from this optimal value is at most 20\%.

\begin{figure}[t!]
\begin{center}
\includegraphics[width=\textwidth]{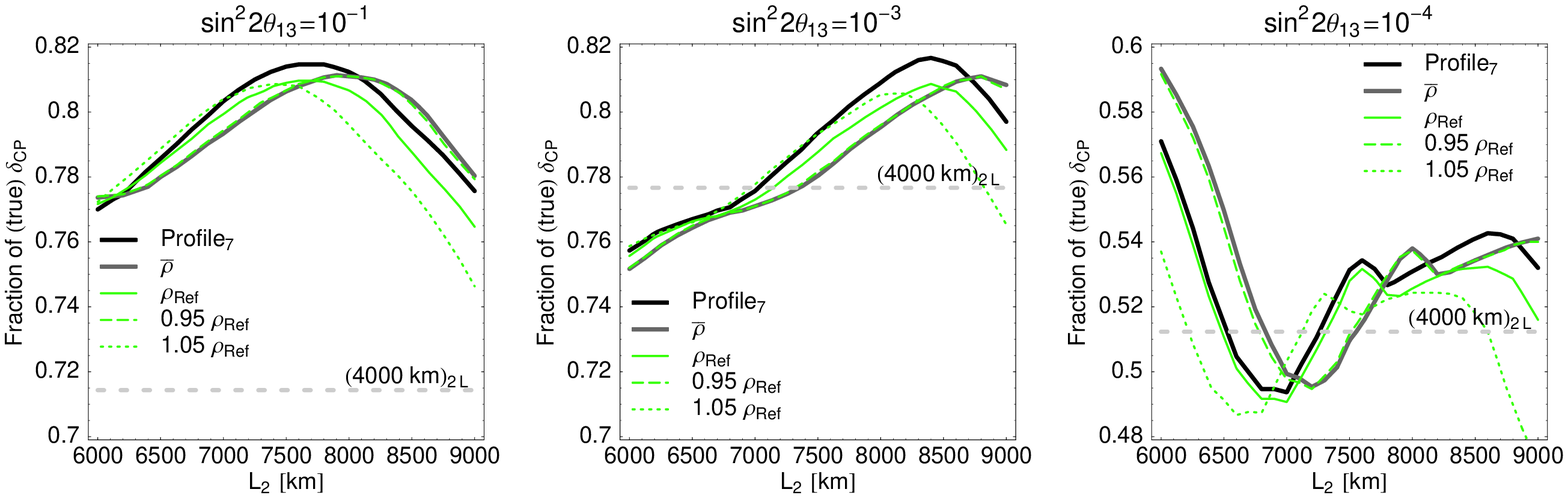}
\end{center}
\mycaption{\label{fig:cp} The ``fraction of (true) $\deltacp$'' for which CP violation can be detected as 
function of $L$  at the $3\sigma$ confidence level. This number represents the fraction of all possible values of $\deltacp$ for which the CP conserving values $\deltacp=0$ and $\pi$ can be excluded. The different panels correspond to different simulated values of $\stheta$, as given in the captions. In all cases, the results are shown in combination with a second baseline $L_1 = 4 \, 000 \, \mathrm{km}$, and the reference curves for placing all detector mass at $L_1$ are shows as well as horizontal lines. The different curves represent our different matter density models as represented in the plot legends. For this figure, a normal mass hierarchy has been assumed.}
\end{figure}

It is well known that  the short baseline plays a major role in enhancing the sensitivity to CP violation.  Thus, in \figu{cp}, we  show the combination of $L_1 = 4 \, 000 \, \mathrm{km}$
and $L_2$ on the horizontal axes for the sensitivity to CP violation, where the long baseline  serve primarily as a degeneracy resolver (note the scale on the vertical axes). For large and medium $\stheta$ (left and middle panels), one can read off that slightly longer baselines compared to the magic baseline are preferred. For example, for large $\stheta$, the optimum would be $L_2 = (7 \, 700 \pm 500)  \, \mathrm{km}$. The reason for this pull to longer baselines is the $\deltacp$ sensitivity returning for baselines longer than the magic baseline, \ie, there is additional statistics on the CP violation measurement itself. This preference of longer baselines becomes more dominant for smaller values of $\stheta$, and is also accompanied by a narrowing of the peak. Only for very small values of $\stheta$ (right panel), does the magic baseline peak at $\sim 7 \, 600 \, \mathrm{km}$ become less pronounced because the measurement is statistics dominated and any baseline with some CP violation sensitivity helps. Again, placing all
detector mass at $4 \, 000 \, \mathrm{km}$ is shown for comparison by the horizontal lines. The 
increase of the physics performance by using the very long baseline is actually best for large $\stheta$, where the correlation with the matter effect affects the short baseline. 
Coming back to our main questions, \figu{cp} illustrates once more that $\rho_{\mathrm{Ref}}$ models
the profile effect very well with respect to the positions of the peaks, while the peaks for  $\bar{\rho}$ are slightly shifted. Note, however, that the absolute performance using the
matter density profile is up to 1\% better in the fraction of $\deltacp$ (profile effect). Misjudging the matter density is only relevant if it turns out to be much higher than anticipated and the chosen baseline is very long (effect up to several per cent). Taking into account all the information from this section, we conclude that the final choice of the very long baseline length may well be determined by the availability of detector locations and the storage ring design rather than from physics, because the combination of two baselines is rather insensitive to the very long baseline length and the specifics of the matter density profile. In particular, a baseline somewhat longer than the magic baseline does not downgrade the  physics
 potential.

\section{Application: $\boldsymbol{\stheta}$ precision measurement}
\label{sec:theta13}

While it is obvious from the phenomenological point of view that a very long baseline is a  robust
correlation and degeneracy resolver, one could conceivably   resolve these degenerate solutions by
different approaches, such as increased statistics, an improved detector, or different oscillation channels.
However, in this section, we will discuss an obvious generic application 
of a very long baseline: the precision of $\theta_{13}$. In general, this measurement depends strongly  on the true value of $\deltacp$~\cite{Huber:2002mx}. Since, to leading order,  there is no $\deltacp$-dependence at the magic baseline, it provides a low risk option for detector placement, in addition to the possibility of improved precision in $\theta_{13}$.

\begin{figure}[t!]
\begin{center}
\includegraphics[height=8cm]{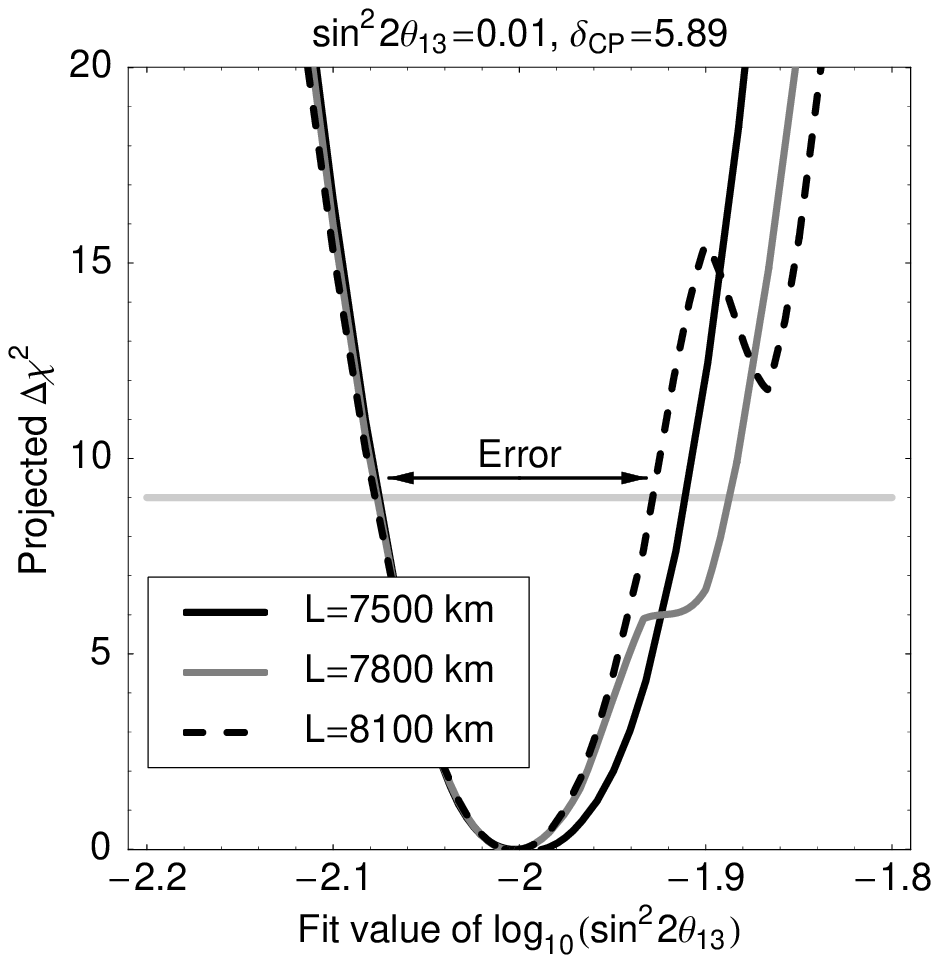}
\end{center}
\mycaption{\label{fig:didtheta13} The (projected) $\Delta \chi^2$ as function of $\log_{10} (\stheta)$
for the true values $\stheta=0.01$ and $\deltacp=5.89$ for different baselines as given in the plot legend.}
\end{figure}

We define the $\stheta$ precision as the full width relative error on $\log_{10} (\stheta)$, \ie,
\begin{equation}
\mathrm{Rel. \, error \, on} \, \log_{10} (\stheta) \equiv \frac{\log_{10} (\stheta) |_{\mathrm{upper}} - \log_{10} (\stheta) |_{\mathrm{lower}}}{\log_{10} (\stheta) |_{\mathrm{true}}} \, ,
\end{equation}
where  ``upper'' and ``lower'' refer to the most upper and lower intersections of the fitted $\Delta \chi^2$ with the line $\Delta \chi^2 = 9$. We do not include the $\mathrm{sgn}(\ldm)$-degeneracy because it is hard to define the $\stheta$ precision including this information. The chosen value $\Delta \chi^2=9$ corresponds to the $3 \sigma$ error for Gaussian errors, which is, strictly speaking, not always given in this case. We illustrate our definition in \figu{didtheta13}, where the (projected) $\Delta \chi^2$ is shown as function of $\log_{10} (\stheta)$ for a specific set of true values. The error is obtained from the intersections with the horizontal line (see arrows). As one can easily see in this figure, the $(\deltacp, \theta_{13})$-degeneracy~\cite{Burguet-Castell:2001ez} is present in some cases under the chosen $\Delta \chi^2$. It is included in the error by our definition. One can also see from this figure that while at the magic baseline the $(\deltacp, \theta_{13})$-degeneracy is non-existent (no $\deltacp$-dependence), it reappears for longer baselines ($L=7 \, 800 \, \mathrm{km}$), but is quickly lifted over the chosen confidence level ($L=8 \, 100 \, \mathrm{km}$). Therefore, we expect jumps in the $\theta_{13}$ precision as a function of baseline wherever this lifting occurs.

\begin{figure}[tp]
\begin{center}
\includegraphics[height=0.73\textheight]{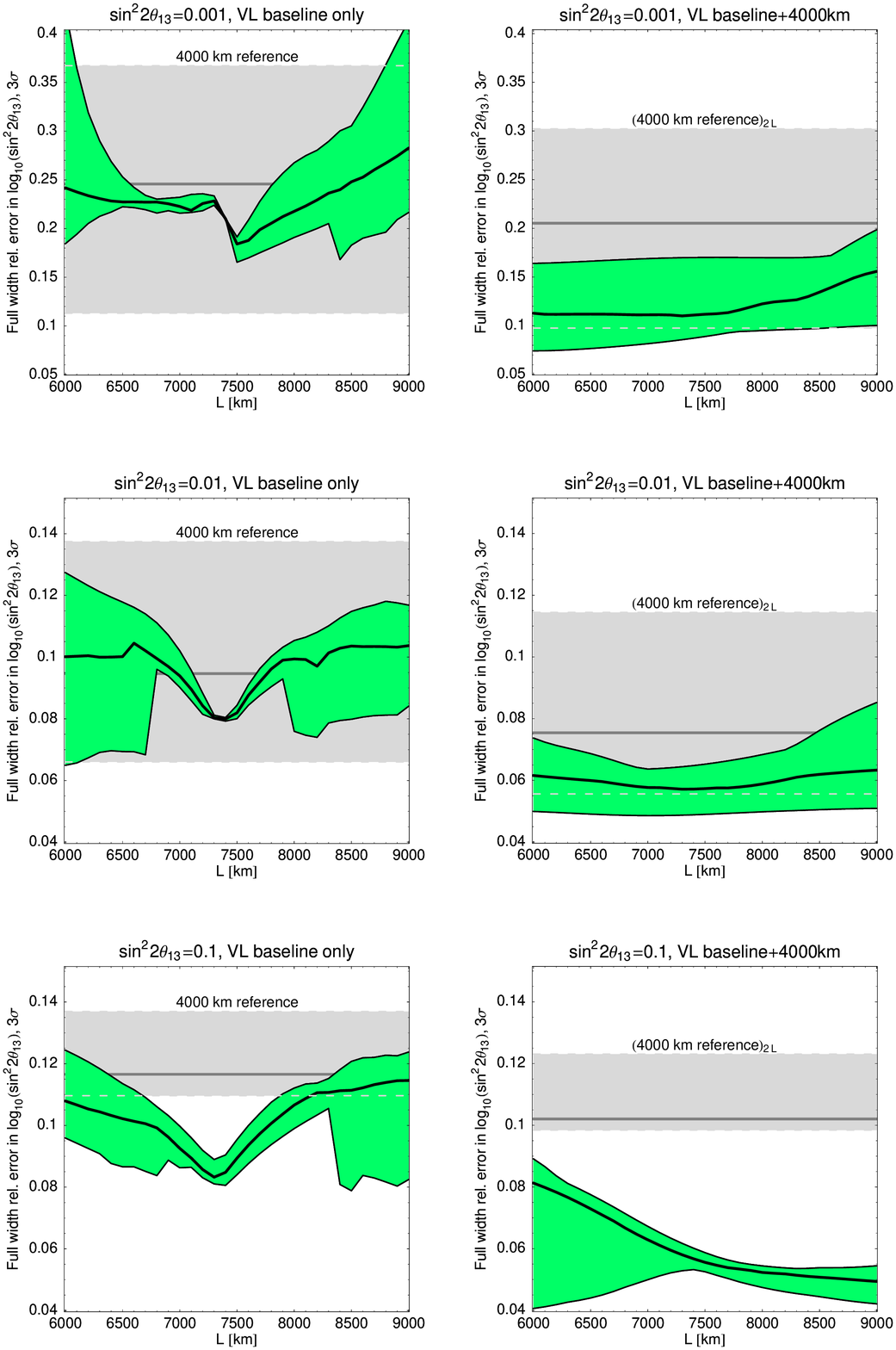}
\end{center}
\mycaption{\label{fig:theta13prec} The precision of $\stheta$ (as defined in the main text) as function of the very long baseline length $L$. The left column corresponds to the very long baseline only, the right column to the combination with a shorter baseline at $4 \, 000 \, \mathrm{km}$. The different rows correspond to different (true) values of $\stheta$ as given in the plot captions. The dependence on the true value of $\deltacp$ is shown by the bands: The upper ends correspond to the worst case $\deltacp$, the lower ends to the best case $\deltacp$, and the thick curves to the ``typical value of $\deltacp$'', \ie, the median of the distribution (the precision will be better in 50\% of all cases, and worse in the other 50\%). The gray horizontal bands represent the short baseline only for reference (worst case, median, and best case). Note that in the right column, double luminosity was used for the reference (corresponding to 
all detector mass at $4 \, 000 \, \mathrm{km}$). In this figure, the reference density $\rho_{\mathrm{Ref}}$ was taken to simulate the matter density profile.}
\end{figure}

We show in \figu{theta13prec} the precision of $\stheta$ as function of the very long baseline length $L$. The left column corresponds to the very long baseline only, the right column to the combination with a shorter baseline at $4 \, 000 \, \mathrm{km}$. The different rows correspond to different (true) values of $\stheta$ as given in the plot captions. The dependence on the true value of $\deltacp$ is shown by the bands: The upper ends correspond to the worst case $\deltacp$, the lower ends to the best case $\deltacp$, and the thick curves to the ``typical value of $\deltacp$'', \ie, the median of the distribution (the precision will be better in 50\% of all cases, and worse in the other 50\%). The gray horizontal bands represent all detector mass at the short baseline.

Let us first of all focus on the left column of \figu{theta13prec} and discuss the $\theta_{13}$ precision for a single very long baseline. There are two aspects which can be inferred from this
figure: The absolute performance and the risk minimization with respect to $\deltacp$. For the
absolute performance, compare, for instance, the thick dark curve for the ``typical'' $\deltacp$ with the reference thick horizontal line for $L = 4 \, 000 \, \mathrm{km}$. One can easily read off that in all cases of $\stheta$ (rows) a baseline of around $6 \, 700 \, \mathrm{km}$ to $7 \, 700 \, \mathrm{km}$ performs significantly better than the short baseline for the typical $\deltacp$. For large or small $\stheta$, this baseline window is even larger. As far as risk minimization is concerned, the worst case performance (upper ends of bands) is almost always better than for the short baseline, whereas the best case performance is only better for large $\stheta$. The reason is the importance of the matter density
uncertainties for large $\stheta$~\cite{Ohlsson:2003ip}, which can be reduced by a clean determination
of $\stheta$ at the very long baseline~\cite{Huber:2006wb}. In addition, one can clearly see that the dependence on $\deltacp$ is minimal at the magic baseline, where the bands become very narrow. 
The jumps in the best case precision come from the lifting of the $(\deltacp, \theta_{13})$-degeneracy
as discussed above. In this case, the precision is determined by a different value of $\deltacp$, leading to the jump.

As for the combination of two baselines, we compare the very long baseline combined with $4 \,000 \, \mathrm{km}$ with all detector mass located at $4 \,000 \, \mathrm{km}$ (gray shaded region).
In all cases, the best case, the worst case, and the median performance are better than for the
short baseline only, which means that the combination of the two baselines is very synergistic.
In particular, the performance for the typical $\deltacp$ is much better for the combination  than for one baseline only.
In addition, there is much less sensitivity to the exact value of the very long baseline length compared to one baseline only. Note that the behavior as function of the baseline depends very much on $\stheta$: For small $\stheta$, shorter baselines are preferred because of better statistics, whereas for large $\stheta$, longer baselines are preferred since matter effects are dominant.\footnote{The reason for the preference of longer baselines may actually be non-trivial: At the magic baseline, the $\sin(\hat{A} \Delta)$-term in \equ{PROBMATTER} changes sign. Therefore, for longer baselines, the role of $\deltacp$ and $-\deltacp$ is exchanged, \ie, the risk with respect to $\deltacp$ is lowered by the combination of the short and long baseline. Note that a similar behavior can be found for the silver channel at the short baseline: 
For maximal 
mixing, the probability is the same as for the golden channel except from $\deltacp \rightarrow -\deltacp$.}

\section{Application: Resolving the $\boldsymbol{(\theta_{23},\pi/2-\theta_{23})}$-degeneracy}
\label{sec:resdeg}

\begin{figure}[t]
\begin{center}
\includegraphics[width=\textwidth]{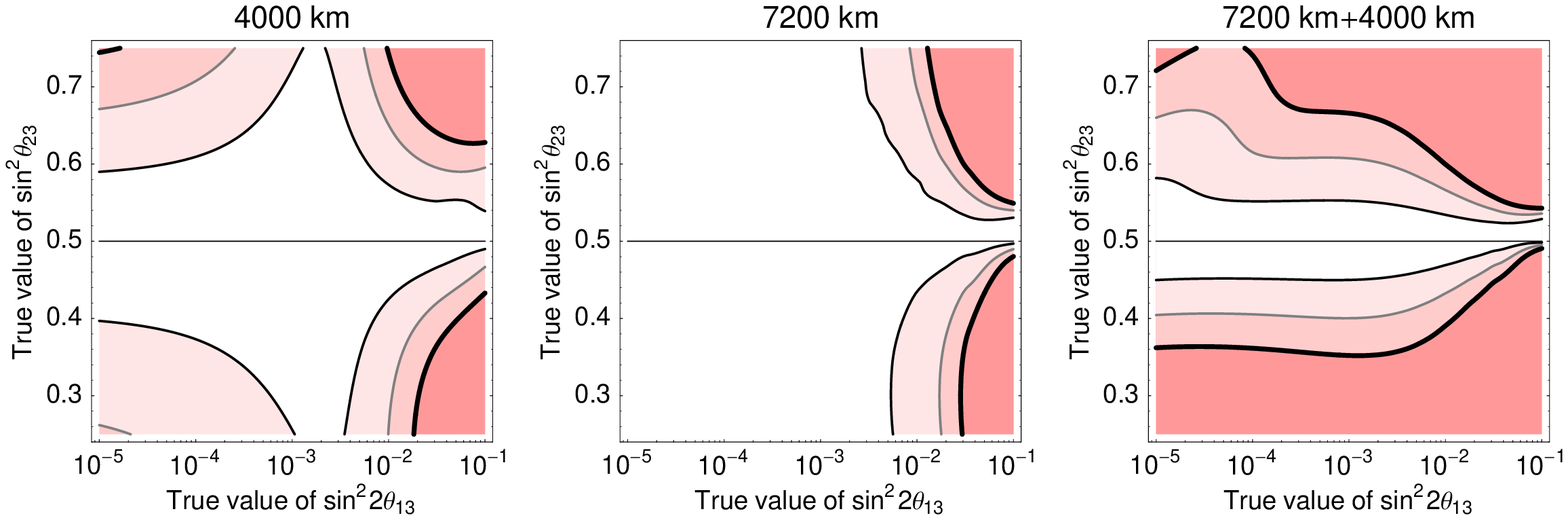}
\end{center}
\mycaption{\label{fig:theta23} Regions where the $(\theta_{23}, \pi/2 - \theta_{23})$-degeneracy can be resolved as function of $\stheta$ and $\sin^2 \theta_{23}$ for different baseline options as given in the plot labels. The different shadings represent the $1\sigma$, $2\sigma$, and $3 \sigma$ confidence level sensitive regions (from light to dark, 1 d.o.f.).
The different panels represent different baseline configurations. Here the true $\deltacp=0$ and a normal mass hierarchy is assumed. Note that this figure does not take into account the mixed degeneracy in $\mathrm{sgn}(\ldm)$ and $(\theta_{23}, \pi/2 - \theta_{23})$ combined.}
\end{figure}

Another interesting application of a very long neutrino factory baseline is the resolution of the $(\theta_{23}, \pi/2 - \theta_{23})$-degeneracy~\cite{Fogli:1996pv}. For example, at the magic baseline, we have from \equ{PROBMATTER}
\begin{eqnarray}
P_{e\mu}^{\mathrm{Magic}} & \simeq & \sin^2 2\theta_{13} \, \sin^2 \theta_{23} \frac{\sin^2[(1- \hat{A}){\Delta}]}{(1-\hat{A})^2} \, , \label{equ:mbgold}
\end{eqnarray}
which should be very sensitive to $\sin^2 \theta_{23}$ and therefore the octant at least for large $\stheta$. However, for small $\stheta$, $P_{e\mu}^{\mathrm{Magic}} \simeq 0$ at this baseline, which means that 
there is no octant sensitivity. However, in the limit $\stheta \rightarrow 0$
\begin{eqnarray}
P_{e\mu}^{\mathrm{L=4 \, 000 \, \mathrm{km}}} & \simeq & \alpha^2 \, \cos^2 \theta_{23}  \sin^2 2\theta_{12} \frac{\sin^2(\hat{A}{\Delta})}{\hat{A}^2}
\end{eqnarray}
at the short baseline, which means that the octant degeneracy can, in principle, be resolved by the $\cos^2 \theta_{23}$-dependence in the solar term at the shorter baseline. Note that this term is suppressed by $\alpha^2 \sim 10^{-3}$, which means that the octant resolution will be considerably worse than for large $\stheta$. For intermediate $\stheta$, all terms in \equ{PROBMATTER} will be present, which means that the sensitivity to the $(\theta_{23}, \pi/2 - \theta_{23})$-degeneracy will be highly affected by correlations with $\deltacp$. We therefore expect that the very long baseline could help to resolve these correlations.

We define that the   $(\theta_{23}, \pi/2 - \theta_{23})$-degeneracy can be considered to have been eliminated if, for a given simulated $\theta_{23}$, no degenerate solution with $\theta_{23}$ in the wrong octant (and the same mass hierarchy) fits the original solution at the chosen confidence level. In order to find the degeneracy, we marginalize over all oscillation parameters in the wrong octant.
We show in \figu{theta23} the sensitivity to exclusion of  the $(\theta_{23}, \pi/2 - \theta_{23})$-degeneracy as a function of $\stheta$ and $\sin^2 \theta_{23}$ for different baseline options. Obviously, for large $\stheta$, the very long baseline has the better sensitivity because it is less affected by correlations with $\deltacp$. For small $\stheta$, the short baseline can still resolve the degeneracy, but with a very poor reach in $\sin^2 \theta_{23}$. For medium $\stheta$, there is no sensitivity for either baseline because the short baseline is affected by correlations, and the long baseline has intrinsically no sensitivity. However, the combination of the two baselines does very well in a wide range of $\stheta$ because the long baseline helps to constrain $\stheta$ very well. Note that we do not show the sensitivity to the mixed (octant and sign) degeneracy in this analysis, which would impose the  additional requirement that the  mass hierarchy be resolved (\cf, \eg, \Ref~\cite{Huber:2005ep}). 
In addition, our results are quantitatively comparable with the ones in \Ref~\cite{Donini:2005db} for the individual baselines. As far as the dependence on $\deltacp$ is concerned, we do not expect a qualitatively interesting behavior, see \Refs~\cite{Huber:2005ep,Donini:2005db}. 

It is possible that
an octant degeneracy resolution may be obtained earlier than from a neutrino factory by atmospheric plus long-baseline data~\cite{Huber:2005ep,Campagne:2006yx}, reactor plus long-baseline data~\cite{Minakata:2002jv,McConnel:2004bd,Hiraide:2006vh}, or long-baseline plus astrophysical data~\cite{Serpico:2005bs,Winter:2006ce}.\footnote{See also \Ref~\cite{Choubey:2005zy} for atmospheric data alone and large $\theta_{13}$.} For large $\stheta$, a neutrino factory would be very competitive to all of these these methods. For small $\stheta$, a determination from atmospheric (or possibly 
astrophysical) data combined with long-baseline data could come earlier. However, it would require a megaton-size water Cherenkov detector and megawatt-class superbeam upgrade. From a practical point of view, it appears likely  that only one of these combinations (neutrino factory or superbeam upgrade) will be eventually  realized.

Finally, we have examined, as  a different approach, the efficiency of the  silver channel at the magic baseline as a degeneracy resolver. At this baseline, we have (see, \eg, \Ref~\cite{Akhmedov:2004ny})
\begin{eqnarray}
P_{e\tau}^{\mathrm{Magic}} & \simeq & \sin^2 2\theta_{13} \, \cos^2 \theta_{23} \frac{\sin^2[(1- \hat{A}){\Delta}]}{(1-\hat{A})^2} \ .
\end{eqnarray}
Therefore, using \equ{mbgold}, the ratio $P_{e \mu}/P_{e \tau} = \tan^2 \theta_{23}$ should determine the
octant degeneracy without being spoilt by any correlations and degeneracies.
We have tested the silver channel for the implementation in \Refs~\cite{Autiero:2003fu,Huber:2006wb} with a $10 \, \mathrm{kt}$ emulsion cloud
chamber and 4~yr of $\nu_\tau$ appearance, and we have 
not found a significant contribution to the two baseline
combination. However, the silver channel does help somewhat for the very long
baseline alone for intermediate $\stheta$. The reason for the marginal contribution is the very low event rate in the silver channel at this baseline, \ie, low statistics (about 176 signal events for $\stheta=0.1$).

\section{Application: Matter density measurement}
\label{sec:matterdensity}
As mentioned earlier, the sensitivity to matter density changes at the magic baseline is high.
Moreover, the fact that matter density uncertainties affect the extraction of $\stheta$ and $\deltacp$
for large $\stheta$ is well known (see, \eg, \Ref~\cite{Ohlsson:2003ip}). 
 This implies that 
one can extract some information on the matter density as well. Such neutrino oscillation tomography
using a neutrino beam has, for example, been studied in \Refs~\cite{Ohlsson:2001fy,Ohlsson:2001ck,Winter:2005we,Minakata:2006am}. In this section, we discuss the
use of neutrino data for potential geophysics applications as a by-product of a very long neutrino factory baseline. We stress however that the primary purpose of such a baseline remains its  use as a degeneracy resolver for neutrino oscillation physics, and the geophysics would be 
a nice addition coming for free. However, if there are different alternative detector locations,
the geophysics along the neutrino factory baseline might have some impact on the choice of the location.

We follow the treatment of the matter density measurement in \Ref~\cite{Winter:2005we} described in \Sec~\ref{sec:methods}, \ie, we use the information from both the short ($L=4 \, 000 \, \mathrm{km}$) and 
the very long baseline 
to reduce the impact of correlations. For the short baseline, we use the mean density in \equ{rhoavg} and allow for a 5\% matter density uncertainty. For the long baseline, we will use several models for the density measurement as described below. Note that all of the oscillation parameters are  marginalized over, \ie, their  uncertainties  are taken into account.

\begin{figure}[t!]
\begin{center}
\includegraphics[width=0.5\textwidth]{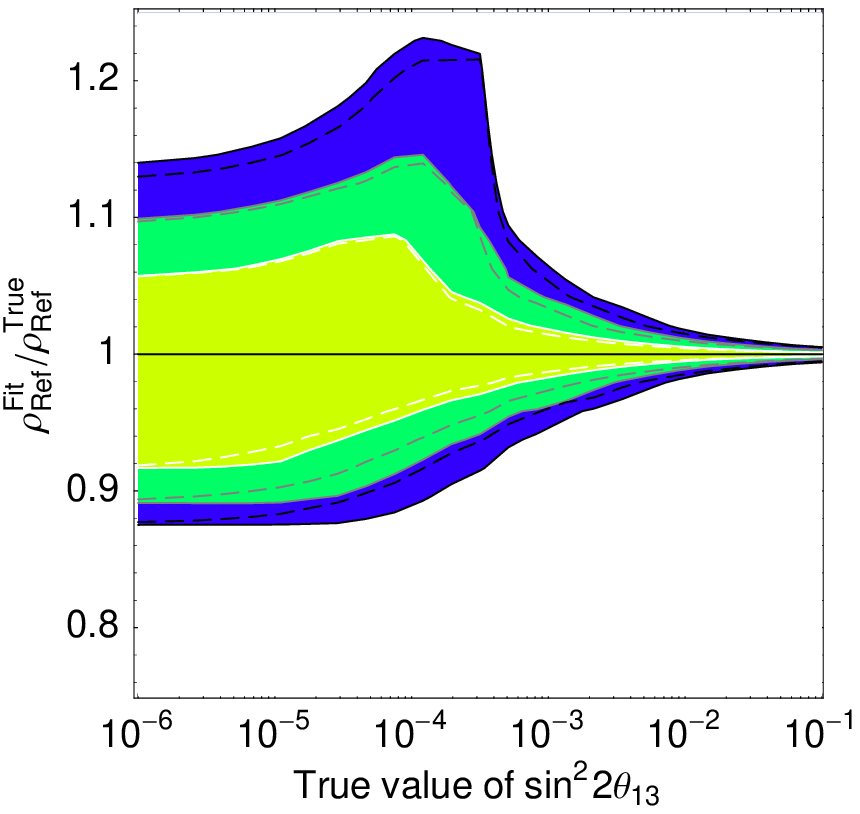}
\end{center}
\mycaption{\label{fig:denstheta13} Relative precision of the density measurement of $\rho_{\mathrm{Ref}}$ as function of
$\stheta$ at the $1\sigma$, $2 \sigma$, and $3 \sigma$ confidence levels (contours from light to dark).
For the baseline, we have chosen $L=7 \, 500 \, \mathrm{km}$ in combination with $L= 4 \, 000 \, \mathrm{km}$ used to measure the oscillation parameters, as well as we have assumed a normal mass hierarchy and $\deltacp=0$. The dashed curves correspond to not taking into account the correlations with the oscillation parameters, which means that the shown combination is close to optimal.}
\end{figure}

\begin{table}[t!]
\begin{center}
\begin{tabular}{lrrrrrr}
\hline
& \multicolumn{3}{c}{Measurement of $\rho_{\mathrm{Ref}}$} & \multicolumn{3}{c}{Measurement of $\rho_{\mathrm{LM}}$} \\
True value of $\stheta$ & $1 \sigma$ & $2 \sigma$ & $3 \sigma$ & $1 \sigma$ & $2 \sigma$ & $3 \sigma$ \\
\hline
\\[-0.3cm]
$\stheta=0.1$ & $_{-0.0024}^{+0.0024}$ & $_{-0.0047}^{+0.0047}$ & $_{-0.0071}^{+0.0071}$ & $_{-0.0025}^{+0.0025}$  & $_{-0.0050}^{+0.0050}$  & $_{-0.0076}^{+0.0075}$  \\[0.1cm]
$\stheta=0.01$ & $_{-0.0065}^{+0.0065}$ & $_{-0.013}^{+0.013}$ & $_{-0.019}^{+0.020}$ & $_{-0.0068}^{+0.0069}$  & $_{-0.014}^{+0.014}$  & $_{-0.020}^{+0.021}$  \\[0.1cm]
$\stheta=0.001$ & $_{-0.019}^{+0.021}$ & $_{-0.038}^{+0.043}$ & $_{-0.056}^{+0.067}$ & $_{-0.020}^{+0.021}$  & $_{-0.040}^{+0.043}$  & $_{-0.059}^{+0.068}$  \\[0.1cm]
\hline
\end{tabular}
\end{center}
\mycaption{\label{tab:prec} Relative precision of the measurement of $\rho_{\mathrm{Ref}}$ (constant density along baseline) and $\rho_{\mathrm{LM}}$ (lower mantle part of the baseline only) at different confidence levels and for different values of $\stheta$. The same parameters as in \figu{denstheta13} are used. Note that for the $\rho_{\mathrm{LM}}$ precision in this table, the upper mantle density is assumed to be known/fixed. This constraint is studied in \figu{rhocorr}. }
\end{table}

As a first model, let us assume that we measure the constant density along the baseline modeled by $\rho_{\mathrm{Ref}}$ (\cf, \Sec~\ref{sec:model}). The precision of this measurement is shown in \figu{denstheta13} as function of $\stheta$ (and in \Tab~\ref{tab:prec} for specific values of $\stheta$). For large $\stheta$, where the $\deltacp$-terms in \equ{PROBMATTER} are small relative to the first term, the measurement is very precise and dominated by a combination of resonance peak position and probability suppression at high energies, \ie, the characteristic spectral dependence of the matter effect (\cf, \Ref~\cite{Minakata:2006am}). For small $\stheta$, however, the measurement is dominated by the $\alpha^2$-term in \equ{PROBMATTER}, which does not have the resonance information and therefore has a lower precision. 
For medium $\stheta$, the $\deltacp$-terms act as a background to the density extraction, which means that the signal is least clean and the performance is worst there. Note that at the magic baseline only the $\stheta$-term is present by definition. However, the suppression of the other terms by $\sin (\hat{A} \Delta ) =0 $ depends on our knowledge of the matter density, which means that any deviations in the matter density make these terms return. 

As an alternative model, one can also think about measuring the density along the baseline part in a depth $d \gtrsim 670 \, \mathrm{km}$ penetrating the lower mantle of the Earth (\cf, \figu{densmapping}, left, where the inner density jumps occur). We call this density the lower mantle mean density $\rho_{\mathrm{LM}}$ along our baseline, which corresponds to the top density in Profile$_7$. If we assume that we know the upper mantle density profile exactly, we obtain the precisions in \Tab~\ref{tab:prec}, where we also compare them to the measurement of $\rho_{\mathrm{Ref}}$. For large $\stheta$, we find that one can measure the density up to 0.25\% at $1 \sigma$ or 0.75\% at $3 \sigma$. These precisions are very competitive to geophysics and could be used to discriminate among different seismic models.

\begin{figure}[t!]
\begin{center}
\includegraphics[width=\textwidth]{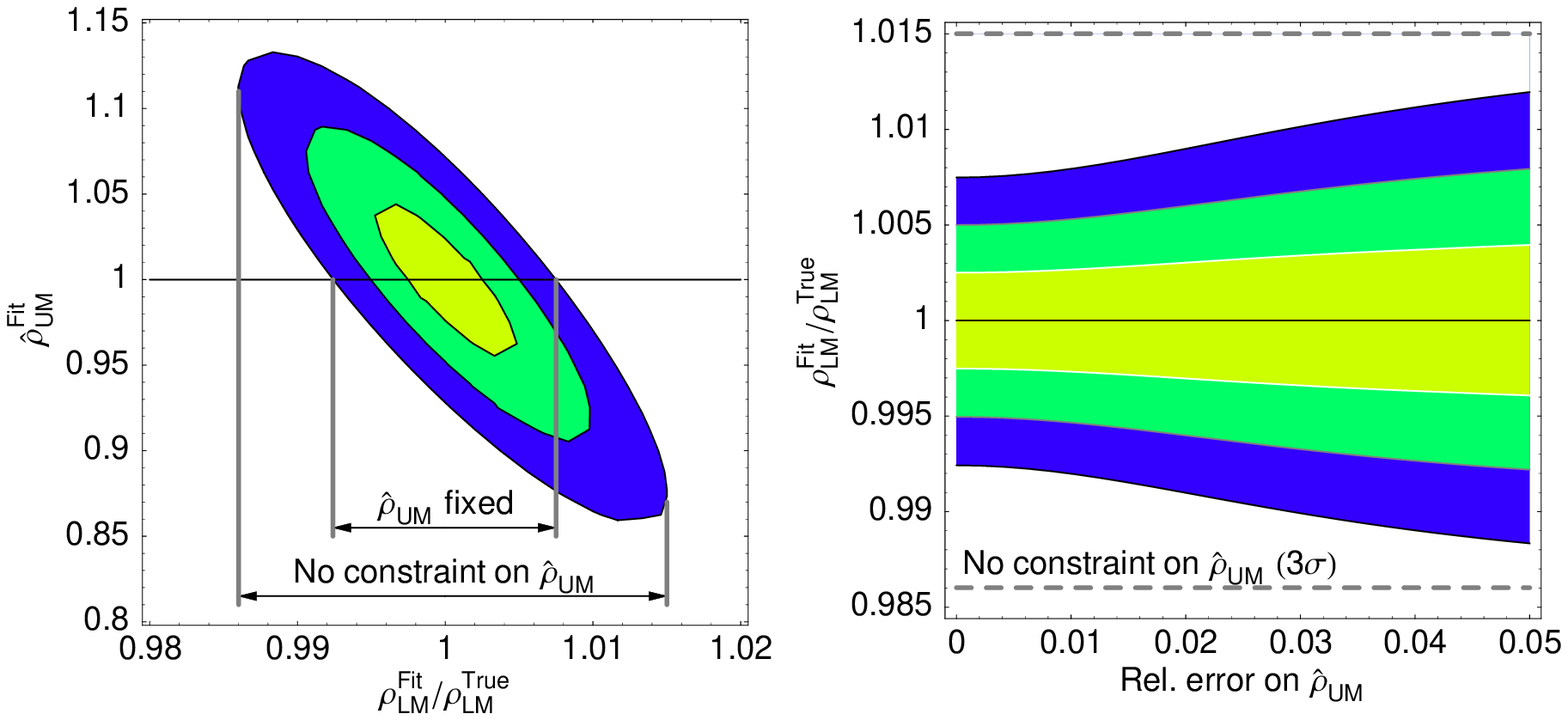}
\end{center}
\mycaption{\label{fig:rhocorr} Left plot: Correlation between lower mantle density $\rho_{\mathrm{LM}}$ and normalization of the upper mantle/crust density $\hat{\rho}_{\mathrm{UM}}$. The arrows indicate the error on  $\rho_{\mathrm{LM}}$ for no constraint on $\hat{\rho}_{\mathrm{UM}}$ and a very precise knowledge on $\hat{\rho}_{\mathrm{UM}}$. Right plot: Precision of $\rho_{\mathrm{LM}}$ as function of relative external error on $\hat{\rho}_{\mathrm{UM}}$. The dashed lines refer to the no constraint limit. Both plots show the $1\sigma$, $2 \sigma$, and $3 \sigma$ errors (from light to dark, 1 d.o.f.) for $\stheta=0.1$.
Here we have chosen $L = 7 \, 500 \, \mathrm{km}$.}
\end{figure}

For a slightly more realistic modeling of the $\rho_{\mathrm{LM}}$ measurement, the obtainable precision will certainly depend on the local knowledge of the upper mantle (and crust) density close to source and detector. This is illustrated by the depth curve in \figu{densmapping}, left, which relatively quickly goes into the lower mantle section. Assuming that \equ{rhoavg} describes the density measurement to first order, we suspect that the upper and lower mantle densities are highly correlated, \ie, for instance, a lower density in the upper mantle can partly be compensated by a higher density in the lower mantle. Therefore, we perform a combined fit and discuss the precision of $\rho_{\mathrm{LM}}$ as function of the error  in  the upper mantle (and crust) density. Note that we only use two parameters in this model: The density $\rho_{\mathrm{LM}}$, which we want to measure, and an overall density normalization $\hat{\rho}_{\mathrm{UM}}$ for the profile in the upper mantle, which we will marginalize over.

The result of this analysis can be found in \figu{rhocorr}. In the left plot, the correlation between $\rho_{\mathrm{LM}}$ and $\hat{\rho}_{\mathrm{UM}}$ is shown. Because the baseline only runs a short distance in the upper mantle, the upper mantle density normalization needs to deviate substantially from unity to cause a major effect in $\rho_{\mathrm{LM}}$ (\cf, scales on axes). The arrows indicate the error on  $\rho_{\mathrm{LM}}$ for no constraint on $\hat{\rho}_{\mathrm{UM}}$ and a very precise knowledge on $\hat{\rho}_{\mathrm{UM}}$. In the right plot of \figu{rhocorr}, we impose an external (Gaussian) $1\sigma$-error on $\hat{\rho}_{\mathrm{UM}}$ which represents the precision we believe in the upper mantle density close to the chosen source and detector locations. We show the precision of $\rho_{\mathrm{LM}}$ as function of this relative external error on $\hat{\rho}_{\mathrm{UM}}$. As one can easily read off this figure, the impact of the upper mantle-lower mantle density correlation will be very small as long as one knows the upper mantle density at the chosen location better than about 2\%, and even for 5\%, representing the worst case, the deviation from the best case is not very strong.

We have also tested the dependence of the measurement on the baseline and we have not found a significant change in the precision as long as $L \gtrsim 7 \, 000 \, \mathrm{km}$. However, not surprisingly, there is a slight optimum at the magic baseline for medium $\stheta$, because the $\deltacp$-dependence is suppressed there. In addition, we have tested the improved neutrino factory from \Ref~\cite{Huber:2006wb} and have found a considerably higher precision if the muon energy is not lowered to $20 \, \mathrm{GeV}$. The reason is that  the precision part comes from the high energy part of the spectrum. Furthermore, we do not expect a quantitatively very different result for different values of $\deltacp$ or the mass hierarchy in the high precision limit, because the $\deltacp$-dependence is of secondary importance for large $\stheta$, and we have considered symmetric operation of  the neutrino factory symmetric with respect to  neutrinos and antineutrinos. For details of the parameter dependencies, see also \Ref~\cite{Minakata:2006am}. Note that compared to \Ref~\cite{Minakata:2006am}, we find slightly higher precisions for large $\stheta$, because we use the full spectral information (in 43 energy bins) and the knowledge from the short baseline.

\begin{figure}[t!]
\begin{center}
\includegraphics[width=\textwidth]{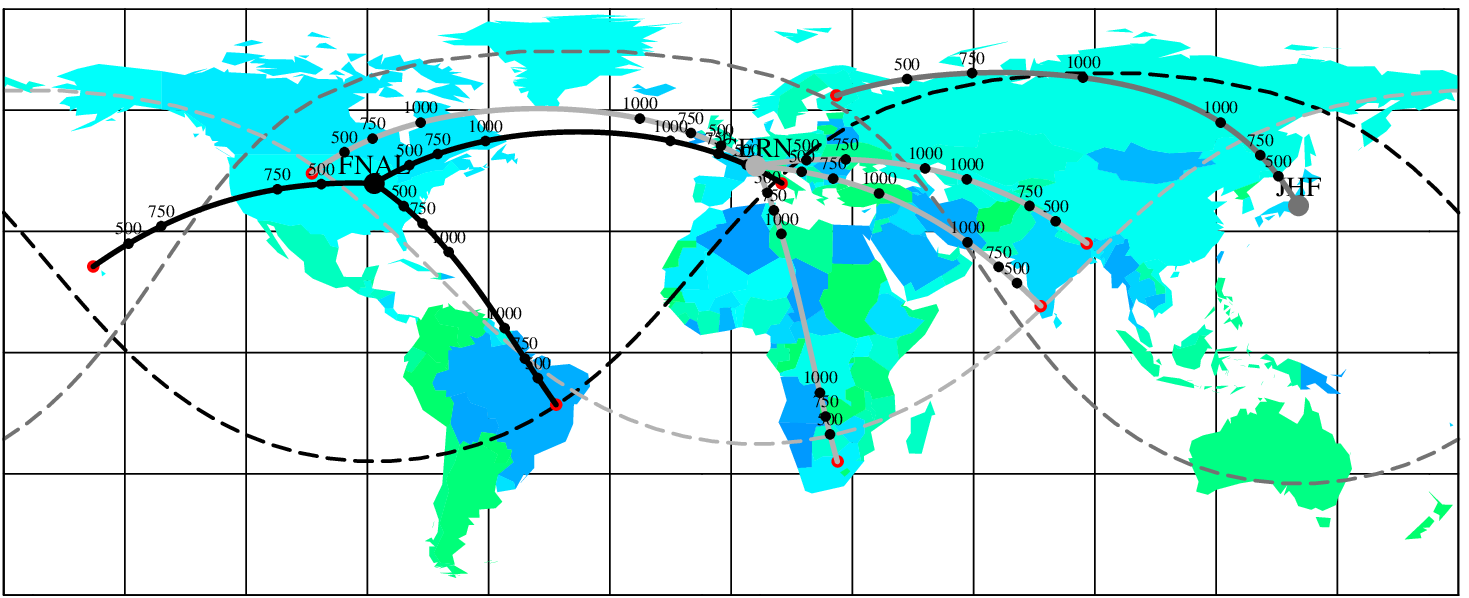}
\end{center}
\mycaption{\label{fig:baselines} Different major potential neutrino factory host laboratories,
all corresponding detector locations at $L = 7 \, 200 \, \mathrm{km}$ (dashed curves, corresponding colors), and several chosen baselines to specific potential detector locations (red points/end points of curves) projected onto the Earth's surface (solid curves). Along these baselines the depth is given in kilometers at several selected points.}
\end{figure}

As far as the geophysical interpretation and the selection of detector locations are concerned, we show in \figu{baselines} different major potential neutrino factory host laboratories and several chosen baselines to specific potential detector locations (red points/end points of curves) projected onto the Earth's
 surface (solid curves). Along these baselines the depth is given in kilometers at several selected points. 
Comparing \figu{baselines} with seismic wave reconstructions (see, \eg, \Ref~\cite{Pana}), one can read off that the obtainable precision on $\rho_{\mathrm{LM}}$ is sufficient for a $3\sigma$ discrimination of different seismic wave reconstructions at specific places, and possibly sufficient for the discrimination of different geophysical hypotheses. However, we leave the drawing of stronger conclusions based on these results to experts in geophysics.

\section{Physics case for a very long baseline $\boldsymbol{6 \, 000 \, \mathrm{km} \lesssim L \lesssim 9 \, 000 \, \mathrm{km}}$  }
\label{sec:physics}

In this section, we summarize the results from this and earlier studies to 
qualitatively review the physics case for a very long neutrino factory baseline. We demonstrate that
such a baseline is extremely versatile over the full allowed range of values of $\stheta$, even if the specific advantages it has differ for low, medium and high values of this parameter. Note that we focus on the baseline window $6 \, 000 \, \mathrm{km} \lesssim L \lesssim 9 \, 000 \, \mathrm{km}$ used for this study.

\subsubsection*{All allowed values of  $\boldsymbol{\stheta}$}

A very long baseline helps for the precision measurements of $\ldm$ and especially $\theta_{23}$, where, as a general rule of thumb, the longer the baseline, the better~\cite{Huber:2006wb}. 
Furthermore, as illustrated in \Sec~\ref{sec:resdeg}, the ability of a shorter baseline to resolve the $(\theta_{23},\pi/2- \theta_{23})$-degeneracy is significantly improved for all values of $\stheta$.  In addition, the very long baseline improves the $\stheta$ precision measurement in all practical cases when $\stheta$ is large enough such that this measurement is relevant. It also minimizes  the uncertainty stemming from  the unknown value of $\deltacp$ (\cf, \Sec~\ref{sec:theta13}).

\subsubsection*{Large $\boldsymbol{\stheta \gtrsim 10^{-2}}$}

For large $\stheta$, the very long baseline helps to reduce the impact of the matter density uncertainties by increasing the accuracy in  the measurement of $\stheta$, which means that the CP violation potential of the short baseline becomes significantly enhanced~\cite{Huber:2006wb}. Note that in this case, an improved knowledge of the  matter
 density profile
 becomes unimportant. Therefore, the very long baseline could be one of the key elements
in establishing the physics case for a neutrino factory for large $\stheta$. In fact, the very long baseline can even measure the matter density along its length to an extremely high precision competitive with geophysical techniques, which may provide additional information on the Earth's lower mantle (\cf, \Sec~\ref{sec:matterdensity} and \Ref~\cite{Minakata:2006am}). 

\subsubsection*{Medium $\boldsymbol{ 10^{-3} \lesssim \stheta \lesssim 10^{-2}}$}

In this scenario, the very long baseline guarantees the mass hierarchy sensitivity for all values of  $\deltacp$, and it resolves correlation and degeneracies affecting the CP violation performance~\cite{Huber:2006wb}. In addition, the $\deltacp$ precision measurement, which strongly depends on the true value of $\deltacp$ itself for the short baseline only, becomes a low risk endeavor~\cite{Huber:2004gg}.
Furthermore, there may still be valuable constraints on the matter density profile (\cf, \Sec~\ref{sec:matterdensity} and \Ref~\cite{Minakata:2006am}).

\subsubsection*{Small $\boldsymbol{10^{-4} \lesssim \stheta \lesssim 10^{-3}}$}

In this case, the main usefulness  of the very long baseline lies in improving  the $\stheta$ and mass hierarchy sensitivities~\cite{Huber:2003ak,Huber:2006wb}. In particular, the mass hierarchy discovery is guaranteed for all values of $\deltacp$, and the CP violation discovery will be possible for at least 50\% of all values of $\deltacp$~\cite{Huber:2006wb}.

\subsubsection*{``Zero'' $\boldsymbol{\stheta ( \ll 10^{-4})}$}

Even if $\stheta$ is far below the limit of a neutrino factory
and $\deltacp$ becomes unmeasurable, there are several interesting measurements which can only be done with a very long baseline . For example, the MSW effect in Earth matter could still be verified at a very high confidence level because the solar appearance term is large enough~\cite{Winter:2004mt}. 
This term would also allow for some information on the octant (see above).
In addition, the disappearance data could be used  towards a possible future measurement of the neutrino mass hierarchy~\cite{deGouvea:2005mi}. 
Finally, this case may be very interesting from the theoretical point of view since it points towards  either an exact or softly broken symmetry.

\section{Summary and conclusions}

We have studied the physics and applications of a very long neutrino factory baseline. Our work was primarily 
motivated by the magic baseline ($L \sim 7 \, 500 \, \mathrm{km}$)
 where correlations and degeneracies naturally disappear. One of the important questions has been the modeling of the matter density
profile, because the  density is crucial for the determination of the magic baseline length.
We have found that the optimal constant density describing the physics at such a baseline is about 5\% higher than the average matter density at this baseline. Note that this optimal density also describes profile effects with sufficient accuracy. This implies that the magic baseline is significantly shorter than one may naively infer, a conclusion that has significant implications for decisions on future detector locations.

Furthermore, we have have re-investigated the baseline optimization for different matter profile assumptions. While a single baseline optimization (magic baseline only) is very sensitive to errors in the assumed   density profile, we have demonstrated that the combination with a shorter baseline $L \simeq 4 \, 000 \, \mathrm{km}$ allows for a much larger baseline window for  detector placement. In particular, for CP violation measurements, much longer baselines $L \sim 7 \, 500$ to $8 \, 500$~km could even be preferable, or, at least, do not harm.

As far as potential applications of a very long baseline are concerned, the primary purpose remains degeneracy resolution. Beyond that, we have demonstrated other interesting applications. For example, we have shown that such a baseline allows for a low risk (with respect to $\deltacp$) $\stheta$ precision measurement, which is, in the worst case, as well as on the average, significantly better than for a short baseline only. In addition, we have illustrated the importance of the very long baseline for the octant degeneracy resolution for intermediate $10^{-4} \lesssim \stheta \lesssim 10^{-2}$. Finally, we have demonstrated that one can also extract the matter density along the baseline with a precision of about $0.24\%$ ($1\sigma$). If one only wants to extract the average lower mantle density of the Earth along the baseline, we still find a precision of about $0.4\%$ for a 5\% matter density uncertainty in the upper mantle (and crust). 

 The physics case for a very long neutrino factory baseline has been summarized in  \Sec~\ref{sec:physics}. Depending on $\stheta$, this baseline has different applications and advantages, but it implies rich possibilities for  all scenarios. Therefore, we conclude that a very long neutrino factory
baseline would be an extremely versatile tool for a broad spectrum of  physics scenarios, and strongly warrants inclusion in the planning of a future neutrino factory complex.

\subsubsection*{Acknowledgments}

We would like to thank Evgeny Akhmedov, Hisakazu Minakata, and the conveners and
contributors of the ISS study for useful discussions.
In addition, WW would like to acknowledge support from the W.~M.~Keck Foundation
through a grant-in-aid to the Institute for Advanced Study, and
through NSF grant PHY-0503584 to the Institute for Advanced Study, where parts of this
work have been carried out, as well as funding by the Emmy Noether program of Deutsche
Forschungsgemeinschaft. RG also thanks the Institute for Advanced Study, Princeton, for hospitality.

\begin{appendix}

\section{How much detector mass is needed at the very long baseline?}
\label{app:detmass}

\begin{figure}[t!]
\begin{center}
\includegraphics[width=0.5\textwidth]{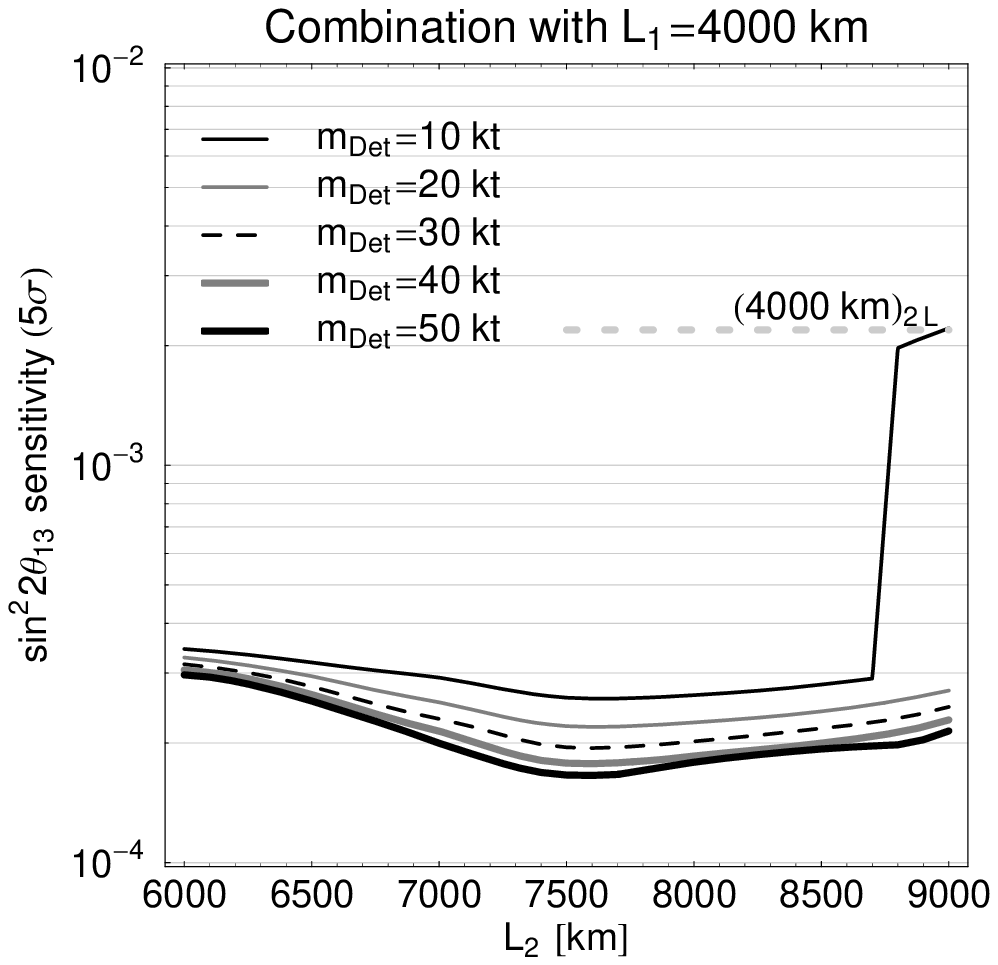}
\end{center}
\mycaption{\label{fig:theta13lumidep} Same as \figu{theta13}, right, for different detector masses 
at the very long baseline $L_2$ as given in the plot legend (the detector mass at $L_1=4 \, 000 \, \mathrm{km}$ is fixed to $50 \, \mathrm{kt}$). For the matter density profile, the reference
density $\rho_{\mathrm{Ref}}$ is used.}
\end{figure}

In order to test the impact of detector mass, we show as one example in \figu{theta13lumidep} the dependence of the $\stheta$ sensitivity on the very long baseline detector mass for the combination with the shorter baseline. One can easily see that $m_{\mathrm{Det}} \gtrsim 20 \, \mathrm{kt}$ at the very long baseline is a safe choice, because smaller detector masses lead to a too large loss of events for baselines  $L_2 \gtrsim 8 \,5 00 \, \mathrm{km}$.

\end{appendix}

\end{document}